\documentclass[11pt]{amsart}
\numberwithin{equation}{section}
\numberwithin{figure}{section}
\numberwithin{table}{section}

\usepackage{stmaryrd}
\usepackage{macros}

\title{The Holomorphic-Topological Poisson Sigma Model and Quantization of Affine $W$-algebras}

\author{Niklas Garner${}^1$\orcidlink{0000-0002-4646-2599}}
\address{${}^1$Mathematical Institute\\ University of Oxford\\ Oxford, OX2 6GG, UK}
\email{\href{mailto:niklas.garner@maths.ox.ac.uk}{niklas.garner@maths.ox.ac.uk}}

\author{Brian R. Williams${}^2$\orcidlink{0000-0001-9641-861X}}
\address{${}^2$Department of Mathematics \& Statistics\\ Boston University\\ 665 Commonwealth Ave, Boston, MA, 02215, USA}
\email{\href{mailto:bwill22@bu.edu}{bwill22@bu.edu}}

\author{Paul-Hermann Balduf${}^{3,4}$\orcidlink{0000-0003-4475-3031}}
\address{${}^3$Department of Combinatorics \& Optimization, University of Waterloo, 200 University Ave. W., Waterloo, ON,  N2L 3G1, Canada}
\address{${}^4$Perimeter Institute, 31 Caroline St. N., Waterloo, ON, N2L 2Y5, Canada}
\email{\href{mailto:paul@paulbalduf.com}{paul@paulbalduf.com}}

\begin{document}
\begin{abstract}
	We construct explicit deformation quantizations of classical principal affine $W$-algebras for every simple Lie algebra $\mathfrak g$ using the holomorphic-topological Poisson sigma model of A. Khan and K. Zeng.
	A linear differential-graded presentation identifies the universal bulk with Chern--Simons theory near a principal nilpotent background, where holomorphic gauge gives a finite, 1-loop exact quantization.
	The quantum oper boundary observables form the principal affine $W$-algebra at level $k-h^\vee\hbar$; we also obtain this level shift directly for affine current algebras.
	For $\mathfrak{sl}_2$, we prove a perturbative quantum equivalence between Chern--Simons theory and three-dimensional chiral gravity, including their oper and Dirichlet boundary conditions, and determine the quantum relation between their parameters.
\end{abstract}

\maketitle
\tableofcontents

\section{Introduction}\label{sec:Introduction}

The Poisson sigma model of \cite{Ikeda:1993fh, Schaller:1994es} is a remarkable two-dimensional topological field theory that provides a physical origin for Kontsevich's deformation quantization of Poisson manifolds \cite{Kontsevich:1997vb, Cattaneo:1999fm}.
In brief, there is a topological field theory $\CT_X$ associated to any Poisson manifold $X$ that admits a natural boundary condition whose classical observables realize its Poisson algebra $A = \CO(X)$ of functions; perturbative quantization of this bulk-boundary system yields a deformation quantization of this Poisson algebra, i.e. an $\hbar$-adic associative algebra $A_\hbar$ together with an isomorphism of Poisson algebras $A \cong A_0 := A_\hbar \otimes_{\C[\![\hbar]\!]} \C$.

Poisson vertex algebras, also called vertex Poisson algebras \cite{DLM, FBZ} or Coisson algebras \cite{Beilinson:2004chiral}, are a chiral generalization of the notion of a Poisson algebra; they can be viewed as the \enquote{classical limit} of a vertex algebra but first appeared in the theory of integrable systems.
An analogue of the Poisson sigma model for Poisson vertex algebras was recently introduced by A. Khan and K. Zeng in the work \cite{Khan:2025rah}; rather than being a fully topological field theory, it is partially holomorphic and hence dubbed the Holomorphic-Topological (HT) Poisson sigma model.
This three-dimensional field theory comes equipped with a natural boundary condition whose classical observables realize the input Poisson vertex algebra.
In the language of \cite{Butson:2016jys}, Poisson algebras and Poisson vertex algebras can be viewed as observables of degenerate field theories, the former being topological and defined on $\R$ and the latter being holomorphic and defined on $\C$, and the ordinary/topological Poisson sigma model and its HT analogue arise from a universal bulk construction.
Just as with the ordinary Poisson sigma model, this higher-dimensional perspective offers a natural quantization procedure for the underlying Poisson vertex algebra by quantizing the combined bulk/boundary system.

Unlike the deformation quantization of Poisson manifolds, it may happen that this quantization procedure is obstructed, i.e. the HT Poisson sigma model $\CT_\CV$ associated to a Poisson vertex algebra $\CV$ may suffer from an anomaly in the presence of this boundary condition.
The classic example of such an obstruction is in the algebra of chiral differential operators on a smooth variety $X$ \cite{GMS00, GMS03, Beilinson:2004chiral}, which is obstructed by the first Pontryagin class of $TX$; see e.g. \cite{Butson:2026vmn} for a recent discussion of the obstruction theory of this quantization problem.
When there is no anomaly, perturbative quantization of the HT Poisson sigma model $\CT_\CV$ is expected to yield a deformation quantization of $\CV$, i.e. an $\hbar$-adic vertex algebra $\CV_\hbar$ together with an isomorphism of Poisson vertex algebras $\CV \cong \CV_0 := \CV_\hbar \otimes_{\C[\![\hbar]\!]} \C$.

In this work we implement this quantization procedure for affine Poisson vertex algebras, classical principal affine $W$-algebras, and the Virasoro Poisson vertex algebra.
The main step is to choose a presentation in which perturbative quantization is tractable.
For higher-rank $W$-algebras, the nonlinear $\lambda$-brackets give a universal bulk theory outside the class of chiral deformations of mixed BF theory, so higher-loop contributions are possible.
We instead use a differential-graded (dg) Poisson vertex algebra with linear $\lambda$-brackets whose cohomology is the classical $W$-algebra: the reduced Drinfel'd--Sokolov complex \cite{dBT94, FBZ}.
Including its differential in the universal bulk construction identifies this bulk theory with Chern--Simons theory expanded about a principal nilpotent background, with oper boundary conditions.
This presentation is a chiral deformation of mixed BF theory.
The holomorphic-gauge construction of \cite{Gwilliam:2019cbp}, with the boundary adaptation described below, gives a finite, 1-loop exact quantization.
Rabinovich's framework for BV theories with boundary \cite{EugeneThesis} supplies the associated factorization algebras of quantum observables.
We compute both the boundary OPEs and the quantum differential, then take cohomology.

The resulting vertex algebra of quantum oper boundary observables is
\be
 W_{k_\hbar}(\fg),\qquad k_\hbar=k-h^\vee\hbar,
\ee
where $h^\vee$ is the dual Coxeter number of $\fg$ (Theorem~\ref{thm:Walgebra}).
Thus the bulk calculation recovers the principal affine $W$-algebra and identifies the affine level entering quantum Drinfel'd--Sokolov reduction.
For Dirichlet boundary conditions, the same method gives the affine current algebra with the same level shift
(Theorem~\ref{thm:1loopaffine}); see \cite{Gwilliam:2019cbp} and \cite[Sec.~7.1]{Costello:2020ndc}.

In rank one, the Virasoro Poisson vertex algebra has two bulk descriptions: the above $\fsl_2$ Chern--Simons theory and three-dimensional chiral gravity.
We prove a classical BV equivalence between them and identify their spaces of perturbative quantizations, both in the bulk and with the corresponding boundary conditions (Theorems~\ref{thm:ClassicalEquivalence}, \ref{thm:cs=g_quantum}, and~\ref{thm:cs=g_boundary}).
For classical parameters $k\neq0$ and $c=-6k$, the quantizations computed in holomorphic gauge correspond under
\be
 c(\hbar)=-6\left(k(\hbar)+\frac{\hbar^2}{k(\hbar)}\right),
 \qquad k(0)=k,\quad c(0)=c.
\ee
This gives a quantum comparison of two presentations of the same classical boundary algebra.

\subsection{Outline}

Section~\ref{sec:CSAffinePVA} develops the method in the affine example.
We fix the mixed BF conventions, explain how bulk quantization determines boundary OPEs and differentials, and compute the affine level shift.

Section~\ref{sec:Walgebras} treats principal affine $W$-algebras.
We describe the reduced Drinfel'd--Sokolov complex \cite{Drinfeld:1985ds}, identify its universal bulk and oper boundary condition, and compute the quantum boundary complex.
Comparing it with quantum Drinfel'd--Sokolov reduction gives the $W$-algebra theorem.

Section~\ref{sec:CSgravity} compares the two bulk realizations of the Virasoro algebra.
After proving the classical equivalence, we compute their quantum boundary algebras and establish the correspondence of quantizations.
The gravitational interpretation connects this comparison to the Chern--Simons description of three-dimensional gravity \cite{Witten:1988hc} and motivates a possible relation to Virasoro TQFT \cite{Collier:2023fwi, Collier:2024mgv}.
That relation is distinct from the perturbative equivalence proved here; see also \cite{Mikhaylov:2017ngi} for a Lagrangian approach to Teichm\"{u}ller TQFT through analytically continued Chern--Simons theory.
The Feynman integrals used throughout are collected in Appendix~\ref{app:integrals}.

\subsection{Outlook}\label{sec:Outlook}

The present work leaves open many directions for future work.

\begin{enumerate}
	\item Extending our analysis to non-principal $W$-algebras is one natural direction for generalization.
	They too will admit a deformation quantization coming from perturbative Chern--Simons theory in the presence of a suitable background determined by the nilpotent element used in the Drinfeld-Sokolov reduction.
	In principle, so long as the nilpotent element admits a good, even gradation, e.g. if $\fg = \fsl_n$, this generalization is merely a matter of algebra.
	\item The equivalence we establish in Section \ref{sec:CSgravity} has a natural extension to higher-rank $\fsl_n$ $W$-algebras and higher-spin gravitational theories.
	This direction is mentioned explicitly in \cite{Khan:2025rah} and we believe that the tools used in the present work could be useful in its pursuit.
	It is worth noting that the $W_{1+\infty}$ algebra itself has linear $\lambda$-brackets and hence should admit a 1-loop exact deformation quantization.
	\item The tools we use to compute the 1-loop Feynman integrals appearing in our deformation quantization can be applied to higher-dimensional analogues of the present setup, e.g. those appearing in \cite{Zeng:2026vry}.
	In ongoing work with P. Bomans and P.-H. Balduf we are applying these techniques to study an analogue of the Chern--Simons/chiral WZW correspondence where a four-dimensional current algebra, cf. \cite{Losev:1995cr, Losev:1996up, Losev1998} and \cite{Faonte2019389, Gwilliam:2018lpo}, lives at the boundary of a five-dimensional holomorphic-topological field theory deforming mixed BF theory.
	\item We expect three-dimensional chiral gravity to arise from the minimal holomorphic--topological twist of three-dimensional $\mathcal N=2$ conformal supergravity.
	The minimal twist of its local superconformal algebra governs deformations of a transverse holomorphic foliation \cite[Sec.~6.1]{Hahner:2024local}, recovering the deformation complex described by $\bmu$.
	The Chern--Simons formulation \cite{Rocek:1986conformal} and superspace constructions \cite{Kuzenko:2012prepotentials, Butter:2013actions} provide possible starting points for deriving the full twisted BV theory.
	Such a derivation would explain the origin of $\bG$ and the $c$-dependent interaction, and identify $c$ in terms of the untwisted supergravity coupling.
	It would also be interesting to interpret the 1-loop central-charge correction in the untwisted theory.
\end{enumerate}

\subsection*{Acknowledgments and Attributions}
The first two authors contributed equally to the main text of the manuscript.
PHB contributed the parametric-space computations of the Feynman integrals outlined in Appendix \ref{app:integrals}.

The authors would like to thank Dylan Butson and Keyou Zeng for useful discussions on chiral quantization, Pieter Bomans for collaboration on related topics and helpful comments on a preliminary version of this paper, and Erik Panzer and Simone Hu for discussions on analytic computation of integrals.
The work of N.G. was previously supported by ERC Consolidator Grant \#864828 \enquote{Algebraic Foundations of Supersymmetric Quantum Field Theory} (SCFTAlg) and is currently supported by the Simons Collaboration on Celestial Holography CH-00001550-1. The work of P.H.B. was  carried out while he was affiliated with the University of Oxford, funded through the Royal Society grants URF{\textbackslash}R1{\textbackslash}201473 and URF{\textbackslash}R{\textbackslash}251041, and also supported by ERC Synergy Grant \#101167287 \enquote{Mathematics of scattering amplitudes} (MaScAmp).

\subsection*{Open Access}
For the purpose of Open Access, the authors have applied a CC BY public copyright license to any Author Accepted Manuscript version arising from this preprint.

\subsection*{Statement of AI Usage}
The authors acknowledge the use of GPT-5.6 Sol and Claude Opus 5.5 for reference searches.
All ideas, arguments, and computations are those of the authors and did not involve the use of generative AI.

\section{Affine algebras from Chern--Simons theory}\label{sec:CSAffinePVA}

This section develops the bulk-boundary quantization procedure in the example of the affine Poisson vertex algebra.
Its universal bulk is perturbative Chern--Simons theory, and the resulting quantization is the vertex-algebra analogue of quantizing functions on $\fg^\vee$ to obtain $U(\fg)$.
We first fix conventions and identify the classical boundary algebra, then explain how the bulk propagators determine the quantum boundary OPEs and differential.
The affine calculation establishes the level shift and the method used for the $W$-algebras in Section~\ref{sec:Walgebras}.

The realization of perturbative Chern--Simons theory that we work with was first introduced in \cite{Aganagic:2017tvx} and studied in greater detail in \cite{Gwilliam:2019cbp}.
Namely, we realize it as a chiral deformation of a mixed BF theory, as defined in Definition \ref{dfn:ChiralDeformation}, cf. Definition 1.2 of \cite{Gwilliam:2019cbp}.
It will be useful to introduce some notation tailored to such theories.
Firstly, we work on a spacetime of the form $\C \times \R$ or $\C \times \R_{\geq 0}$; we often work with local coordinates $(z,\oz)$ on $\C$ and $t$ on $\R$.
On such manifolds, the space of differential forms can be expressed as
\be
	\Omega^n(\C \times \R) = \bigoplus_{p+q+r = n} \Omega^{p,q}(\C) \wh{\otimes} \Omega^r(\R)
\ee
where $\wh{\otimes}$ is the completed projective tensor product.
For $n > 0$ there is the natural subspace
\be
	\bigoplus_{q+r = n-1} \Omega^{1,q}(\C) \wh{\otimes} \Omega^r(\R) \subset \Omega^n(\C \times \R)
\ee
of differential forms in the ideal generated by $\d z$.
We will make use of the quotient
\be
	\bOmega^n := \left\{\begin{aligned}& \Omega^0(\C \times \R) & n = 0\\
		& \frac{\Omega^n(\C \times \R)}{\bigoplus_{q+r = n-1} \Omega^{1,q}(\C) \wh{\otimes} \Omega^r(\R)} & \text{else}
	\end{aligned}\right.~.
\ee
In the above local coordinates, elements of $\bOmega^n$ can be identified with $n$-forms that are not proportional to $\d z$.%
\footnote{Although we do not strictly need to for the present work, we prefer to work with the quotient by differential forms proportional to $\d z$ rather than simply differential forms that are not proportional to $\d z$ as the former generalizes to any 3-manifolds equipped with a transverse holomorphic foliation.} %
As the subspace of forms proportional to $\d z$ is preserved by the exterior derivative, the de Rham differential $\d$ induces a natural map $\d': \bOmega^n \to \bOmega^{n+1}$ making it into a complex $(\bOmega^\bullet, \d')$ analogous to the de Rham complex; in the above local coordinates it takes the form
\be
	\d' = \d \oz \partial_{\oz} + \d t \partial_t
\ee
and hence can be thought of as the sum of the Dolbeault differential on $\C$ and the de Rham differential on $\R$.
We note that $\partial_z$ gives rise to a natural even derivation on the complex $(\bOmega^\bullet, \d')$ that we continue to denote by $\partial_z$.

We will also make use of a slight generalization of the above.
Namely, denote by $K$ the pull back of the canonical line bundle of $\C$ along the projection from $\C \times \R$.
For $j \in \Z$ we can consider the above quotient but with forms valued in the $j$th power of $K$ and use the notation $\bOmega^{(j),\bullet}$.
Locally, elements of this complex transform like forms proportional to $\d z^{\otimes j}$ for $j \geq 0$ and proportional to $\partial_z^{\otimes |j|}$ for $j < 0$.
We place $K$ and its tensor powers in cohomological degree zero: the symbol $\d z$ in a $K$-valued holomorphic-topological form is therefore even, whereas $\d\oz$ and $\d t$ are odd.
Products use the tensor-product Koszul rule, including the ghost degrees of the field components.

\subsection{Chern--Simons theory from mixed BF theory}\label{sec:ClassicalCS}
Let $\fg$ be a simple Lie algebra.
Write $M$ for the spacetime. The fields are
\be
    \bA\in\bOmega^{(0),\bullet}[1]\otimes\fg
    \qquad
    \bB\d z\in\bOmega^{(1),\bullet}\otimes\fg^\vee~.
\ee
Here $[1]$ lowers cohomological degree by one.
The BV pairing is the canonical shifted cotangent pairing between these two summands; each summand is isotropic.
Explicitly, suppose $a_p\in\bOmega^p\otimes\fg$ (of degree $p-1$) and $b_q\in\bOmega^q\otimes\fg^\vee$ (of degree $q$).
On compactly supported inputs, the BV pairing is defined by
\be\label{eq:BVPairingConvention}
\begin{aligned}
    \Omega_{BV}(a_p,\d z\otimes b_q)
    &=(-1)^{pq+1}\int_M\d z~\langle b_q\wedge a_p\rangle~,\\
    \Omega_{BV}(\d z\otimes b_q,a_p)
    &=(-1)^q\int_M\d z~\langle b_q\wedge a_p\rangle~.
\end{aligned}
\ee
Here $\langle-,-\rangle$ is the evaluation pairing $\fg^\vee\otimes\fg\to\C$, extended to differential forms by the wedge product.
The factor $(-1)^{pq}$ in the first line places the $\fg^\vee$-valued form first.
On the left, $\d z$ denotes the frame for the canonical bundle; on the right, the integrands are ordinary exterior forms, integrated with the orientation $\d x~\d y~\d t$ for $z=x+\ii y$.
The pairing is zero unless $p+q=2$, so the field degrees satisfy $(p-1)+q=1$ so that $\Omega_{BV}$ has cohomological degree $-1$.
The factor $(-1)^q$ is the Koszul sign associated to the shift $[1]$; the pairing is graded skew-symmetric:
$\Omega_{BV}(u,v)=-(-1)^{|u||v|}\Omega_{BV}(v,u)$.
With these suspension signs included, we write the pairing in superfield notation as
\be
    \Omega_{BV}=\int\d z~ \langle\delta\bB, \delta\bA \rangle~.
\ee
The action functional is
\begin{align*}
  S &= \int \d z\left\{\langle\bB, \d' \bA + \tfrac{1}{2}[\bA,\bA]\rangle + \frac{\ii k}{4 \pi} \Tr(\bA \partial_z \bA)
  \right\}~ \\
    & = S_{BF} + \int \d z \frac{\ii k}{4 \pi} \op{Tr}(\bA \del_z \bA) 
\end{align*}
where $\Tr$ is the Killing form on $\fg$ normalized to be the trace in the fundamental representation when $\fg = \fsl_N$.
The $k$-independent term is simply the action functional of mixed BF theory.

If we choose a basis $\{T_a\}$ of $\fg$ and denote the commutation relations as
\be
	[T_a, T_b] = f^c_{ab} T_c
\ee
and the normalized Killing form as
\be
	\Tr(T_a T_b) = K_{ab}
\ee
then the above action has the following component form
\be
	S = \int \d z\left\{\bB_a\big(\d' \bA^a + \tfrac{1}{2}f^a_{bc}\bA^b \bA^c\big) + \frac{\ii k K_{ab}}{4 \pi} \bA^a \partial_z \bA^b\right\}
\ee
and corresponding equations of motion encoded in the action of the BV differential
\be
	Q \bA^a = \d' \bA^a + \tfrac{1}{2} f^a_{bc} \bA^b \bA^c \qquad Q \bB_a = \d' \bB_a + f^b_{ac} \bA^c \bB_b + \frac{i k K_{ab}}{2\pi} \partial_z \bA^b~.
\ee
For $k\neq0$, this action is obtained from the ordinary BV description of Chern--Simons theory as follows.
Set
\be
    \CA^a = \bA^a - \d z\,\frac{2\pi K^{ab}}{\ii k} \bB_b~.
\ee
Then
\be
    S_{\mathrm{CS}}=-\frac{\ii k}{4\pi}\int\Tr\left(\CA\d\CA+\tfrac{2}{3}\CA^3\right) .
\ee
In particular, for $k \ne 0$, this deformation of mixed BF theory is classically equivalent to Chern--Simons theory.

\subsection{Affine Poisson vertex algebras}\label{sec:AffinePVA}
One utility of the above rewriting is to manifest the fact that classical Chern--Simons theory can be viewed as a universal bulk theory, as defined by \cite{Butson:2016jys}, for a degenerate classical field theory whose observables give rise to the affine Poisson vertex algebra, as we now review; see Section 4.2 of \cite{Butson:2016jys} for more details or Section 3.1 of \cite{Khan:2025rah} for a complementary discussion.

A degenerate classical BV theory on a surface, equipped with a local $(-1)$-shifted Poisson structure, determines a three-dimensional theory on a collar together with a canonical boundary condition recovering the original theory.
The boundary phase space of the universal bulk is the local higher Poisson center, which carries a $0$-shifted symplectic structure;
the universal bulk is obtained by extending in the topological normal direction.
After passing to holomorphic local operators, the boundary theories in our examples are described by
(dg) Poisson vertex algebras.
Their $\lambda$-brackets express the coisson structure in local coordinates, while the local BV description retains the Dolbeault resolution and factorization structure; see \cite[Sec.~2.6]{Beilinson:2004chiral} and \cite[Secs.~4.3--4.4]{Butson:2016jys}.

When classical Chern--Simons theory is formulated as above then it can be identified with the universal bulk for the affine Poisson vertex algebra $\CV_k(\fg)$, cf. \cite[Sec.~4.2--4.3]{Butson:2016jys} and \cite[Sec.~3.1]{Khan:2025rah}.
This Poisson vertex algebra is freely generated by fields $J_a$ with $\lambda$-bracket
\be
	\{J_a {}_\lambda J_b\} = f^c_{ab} J_c + \lambda k K_{ab}~.
\ee
The resulting universal bulk theory has fields $\bJ_a \d z \in \bOmega^{(1),\bullet}$ together with their canonical conjugates $\boldsymbol{\beta}^a \in \bOmega^{(0),\bullet}[1]$ with BV symplectic form
\be
	\Omega_{BV} = \int \d z ~ \delta \bJ_a \delta \boldsymbol{\beta}^a~.
\ee
and action functional
\be
	S_{\mathrm{univ}} = \int \d z \bigg\{\boldsymbol{\beta}^a \d' \bJ_a + 2 \pi \ii \big(\boldsymbol{\beta}^a \boldsymbol{\beta}^b (\tfrac{1}{2} f^c_{ab}\bJ^c) - \tfrac{1}{2}\boldsymbol{\beta}^a \partial_z \boldsymbol{\beta}^b (k K_{ab}) \big)\bigg\}
\ee
where first term is the standard kinetic term and the terms quadratic in $\boldsymbol{\beta}$ encode the $\lambda$-bracket of the affine Poisson vertex algebra.
That this is equivalent to the above presentation of Chern--Simons theory can be witnessed via the identification
\be
	\boldsymbol{\beta}^a = (2\pi \ii)^{-1} \bA^a \qquad \bJ_a = (2 \pi \ii) \bB_a~.
\ee
The boundary condition giving rise to the affine Poisson vertex algebra $\CV_k(\fg)$ is the \enquote{Dirichlet} boundary condition that sets $\boldsymbol{\beta}^a| = 0$; this is the \enquote{chiral} boundary condition studied in Section 6 of \cite{Gwilliam:2019cbp}.

\subsection{Perturbative quantization and boundary vertex algebras}\label{app:PerturbativeQuantization}
We recall the finite, 1-loop exact quantization of chiral deformations of mixed BF theory from \cite{Gwilliam:2019cbp}, then explain how to extract the boundary vertex algebra.
The same procedure will be used for both the affine and $W$-algebra examples.

\subsubsection{Quantization and propagators}
The underlying free theory has an action functional of the form
\be
	\int \d z~\bB \d' \bA~.
\ee
All the theories we consider in this paper have action functionals of the form
\be
	\int \d z \bigg\{ \bB \d' \bA + \CI(\bB, \bA)\bigg\}~.
\ee

\begin{dfn}\label{dfn:ChiralDeformation}
	The interaction $\CI(\bB, \bA)$ is called \defterm{chiral} if it is a translation-invariant polynomial in the fields $\bB$, $\bA$ satisfying
	\begin{enumerate}
		\item[1)] it is at most linear in $\bB$, i.e. $\CI(\bB, \bA) = \CI_0(\bA) + \bB \CI_1(\bA)$,
		\item[2)] it only involves constant-coefficient differential operators built from $\partial_z$.
	\end{enumerate}
\end{dfn}

As noted in \cite{Gwilliam:2019cbp}, perturbative quantization of any chiral deformation of a mixed BF theory does not admit connected Feynman diagrams beyond 1-loop so long as the gauge fixing respects the $\bA$--$\bB$ splitting.
In fact, Theorem 5.1 of loc. cit. says more: there is a choice of gauge-fixing so that the resulting perturbative quantization is a \emph{finite}, 1-loop exact perturbative quantization.
The gauge-fixing relies on the fact that spacetime is of the form $\C \times \R$ (or $\C \times \R_{\geq 0}$); more generally, it can be defined on any 3-manifold (possibly with boundary) equipped with a transverse holomorphic foliation.
This \emph{holomorphic} gauge-fixing requires
\be
(2 \ol{\partial}^* + \d_t^*) \bA = 0 \qquad (2 \ol{\partial}^* + \d^*) \bB = 0
\ee
where $\ol{\partial}^*$ (resp. $\d_t^*$) is the adjoint of $\ol{\partial}$ (resp. $\d_t$) with respect to the standard Euclidean metric on $\C$ (resp. $\R$).
The propagator for this choice of gauge-fixing takes the form
\be
\begin{aligned}
	\Pi(z,\oz,t; w,\ow,s) & := \langle \bB(z,\oz,t) \bA(w,\ow,s) \rangle\\
	& = \frac{1}{4\pi \ii} \frac{(t-s)\d(\oz - \ow) -2 (\oz - \ow)\d(t-s)}{(|z-w|^2 + (t-s)^2)^{\scriptstyle{\frac{3}{2}}}}
\end{aligned}~.
\ee

When we impose boundary conditions on these fields, the propagator must be modified accordingly.
There are two sorts of boundary conditions we will consider: either $\bB| = 0$ or $\bA| = 0$ which we refer to as \enquote{Neumann} and \enquote{Dirichlet,} respectively.
The relevant propagators $\Pi_N$ and $\Pi_D$ for these two choices take the form%
\footnote{The expression $\Pi(z,\oz,-t; w,\ow,s)$ should be understood as the pullback of $\Pi(z,\oz,t;w,\ow,s)$ along the map $t \to -t$.
Similarly for $\Pi(z,\oz,t; w,\ow,-s)$.}%
\be
	\Pi_N(z,\oz,t; w,\ow,s) = \tfrac{1}{2}\big(\Pi(z,\oz,t;w,\ow,s) - \Pi(z,\oz,-t; w,\ow,s)\big)
\ee
and
\be
	\Pi_D(z,\oz,t; w,\ow,s) = \tfrac{1}{2}\big(\Pi(z,\oz,t;w,\ow,s) - \Pi(z,\oz,t; w,\ow,-s)\big)~.
\ee

\subsubsection{Boundary vertex algebras}\label{sec:BoundaryVertexAlgebras}
We obtain the boundary OPEs and quantum differential from bulk diagrams.
Taking cohomology gives the vertex algebra of boundary observables.

Fix a boundary condition compatible with the interaction and a perturbative BV quantization preserving the
holomorphic--topological structure and satisfying the quantum master equation.
The boundary observables form a complex with quantum differential $Q_{\partial,\hbar}$, whose products are computed by placing observables in disjoint boundary disks and evaluating bulk Feynman diagrams.
Antiholomorphic translations are homotopically trivial, so these products are holomorphic away from coincident points after passing to cohomology.
This is part of the boundary factorization algebra description: bringing the insertion points together encodes the operator
product expansion.

Indeed, using the result of \cite{EugeneThesis} a solution to the quantum master equation on a three-manifold with boundary produces a factorization algebra of quantum observables $\op{Obs}$ on $\C \times \R_{\geq 0}$.
Our aim is to characterize the factorization algebra of \textit{boundary observables} defined by 
\begin{equation}\label{eq:BoundaryObservables}
  \op{Obs}_{\del} = i_{\del}^! \op{Obs} , \quad i_{\del} \colon \C \hookrightarrow \C \times \R_{\geq 0} .
\end{equation}
Here $i^!_{\del}$ is a way of restricting the full factorization algebra to the boundary.
To an open set $U$ on the boundary it is defined to be the limit as $\epsilon \to 0$ of $\Obs(U \times [0,\epsilon))$. 
Since our theories are \textit{topologically normal to the boundary} \cite{EugeneThesis}, there is an equivalence of
factorization algebras $\Obs_{\del} \simeq \pi_* \Obs$, where
$\pi\colon\C\times\R_{\geq0}\to\C$ is the projection $\pi(z,t)=z$.
Thus $(\pi_*\Obs)(U)=\Obs(U\times\R_{\geq0})$.

This factorization algebra on $\C$ satisfies the criteria of \cite[Sec.~5.3]{CG1} and thus defines a vertex algebra.
We now explicitly describe how the resulting OPE can be computed directly in terms of our quantization.
Place two elementary boundary observables $\CO_i(z)$ and $\CO_j(w)$ at distinct points of $\C\times\{0\}$.
The connected contribution to their product is a sum over graphs joining these labelled insertions, with bulk vertices supplied by $\CI$ and an $\bA$--$\bB$ propagator on every edge.
Use $\Pi_D$ or $\Pi_N$ according to the boundary condition on the paired fields; an edge ending at a boundary insertion uses the corresponding restriction of this kernel.
Holomorphic derivatives in a vertex act on its attached propagators or uncontracted fields.
Integrate over the positions of the bulk vertices in the half-space, including the interaction coefficients, graph symmetry factors, and the tensor-product Koszul signs and BV pairing conventions fixed above.
The uncontracted legs represent the output observable; graphs with no such legs contribute multiples of the identity.
For a graph with $E$ propagators, including those incident to the boundary, and $V$ classical interaction vertices, the power of $\hbar$ is $\hbar^{E-V}$.
With two elementary insertion vertices this equals $\hbar^{L+1}$, where $L=E-V-1$ is the loop number.
Since every interaction vertex and every elementary insertion supplies at most one $\bB$-leg, $E\leq V+2$, so these graphs contribute only at tree-level or at one-loop, namely $\hbar$ or $\hbar^2$.

To read off the OPE, expand the uncontracted fields about the second insertion, express the resulting local observables in holomorphic boundary jets, and retain the poles as $z\to w$; the tree-level version of this prescription is described in \cite[Sec.~5.1]{Khan:2025rah}.
Writing the resulting coefficients as $C_{ij}^{(n)}$, the singular OPE and its equivalent $\lambda$-bracket are
\be\label{eq:BoundaryOPEPrescription}
\begin{aligned}
\CO_i(z)\CO_j(w)&\sim\sum_{n\geq0}\frac{C_{ij}^{(n)}(w;\hbar)}{(z-w)^{n+1}},\\
[\CO_i{}_{\lambda}\CO_j]&=\sum_{n\geq0}\frac{\lambda^n}{n!}C_{ij}^{(n)}(\hbar)~.
\end{aligned}
\ee
The coefficients are local, normally-ordered expressions in the boundary generators and their holomorphic derivatives.
The renormalized constant term of the regular part defines the normally-ordered product, while $\partial_z$ generates descendants.
At tree level, the coefficient of $\hbar$ in the $\lambda$-bracket recovers the classical Poisson vertex bracket.
Composite insertions allow additional contractions, so their OPEs need not terminate at order $\hbar^2$ even when the bulk quantization is one-loop exact.

To obtain the boundary vertex algebra one must also compute $Q_{\partial,\hbar}$ and pass to its cohomology \cite[Secs.~2.3--2.4]{Costello:2020ndc}.
Compatibility of the quantum BV differential with the factorization products supplies the identities needed for the OPEs to descend to cohomology.
When these operations admit a strict differential-graded vertex algebra presentation, as in the examples below, $Q_{\partial,\hbar}$ commutes with $\partial_z$ and is a graded derivation of every $n$-product, including the normally-ordered product.
These are quantum products: expressing the differential on composites can itself require normal-ordering corrections.
Thus the three-dimensional description computes both the OPEs and the differential of a two-dimensional presentation, whose cohomology is the boundary vertex algebra.
In Section~\ref{sec:AffineVOA} the surviving currents have trivial differential and the diagrams directly give the affine Kac--Moody OPE.
For the $W$-algebra examples of Section~\ref{sec:Walgebras}, we instead retain the boundary BRST complex and its differential, and extract the $W$-algebra by taking cohomology.

\subsection{Quantization of boundary observables}\label{sec:AffineVOA}

We apply this procedure to the Dirichlet boundary condition for Chern--Simons theory.
The tree-level computation recovers the affine Poisson vertex algebra, as in \cite[Sec.~3.1]{Khan:2025rah}; the one-loop integral determines its quantum level.
The holomorphic-gauge quantization satisfies the quantum master equation by \cite{Gwilliam:2019cbp}.

\begin{thm}\label{thm:1loopaffine}
The holomorphic gauge-fixing of perturbative Chern--Simons theory on $\C\times\R_{\geq0}$ at classical level $k$ with Dirichlet boundary conditions determines a vertex algebra of quantum boundary observables isomorphic to the $\hbar$-adic affine vertex algebra
\be
    V_{k_\hbar}(\fg),\qquad k_\hbar=k-h^\vee\hbar,
\ee
where $h^\vee$ is the dual Coxeter number of $\fg$.
\end{thm}

\begin{proof}
    We compute the singular OPE of the elementary boundary observables $\bB|$.
    Its tree-level terms recover the affine Poisson vertex bracket, and its one-loop term shifts the central coefficient.
	There are three diagrams that contribute to these singular terms, illustrated in Figure \ref{fig:affineFeynman}.
	For the sake of brevity we relegate the explicit computation of these integrals to Appendix \ref{app:AffineOPEs}.
	
	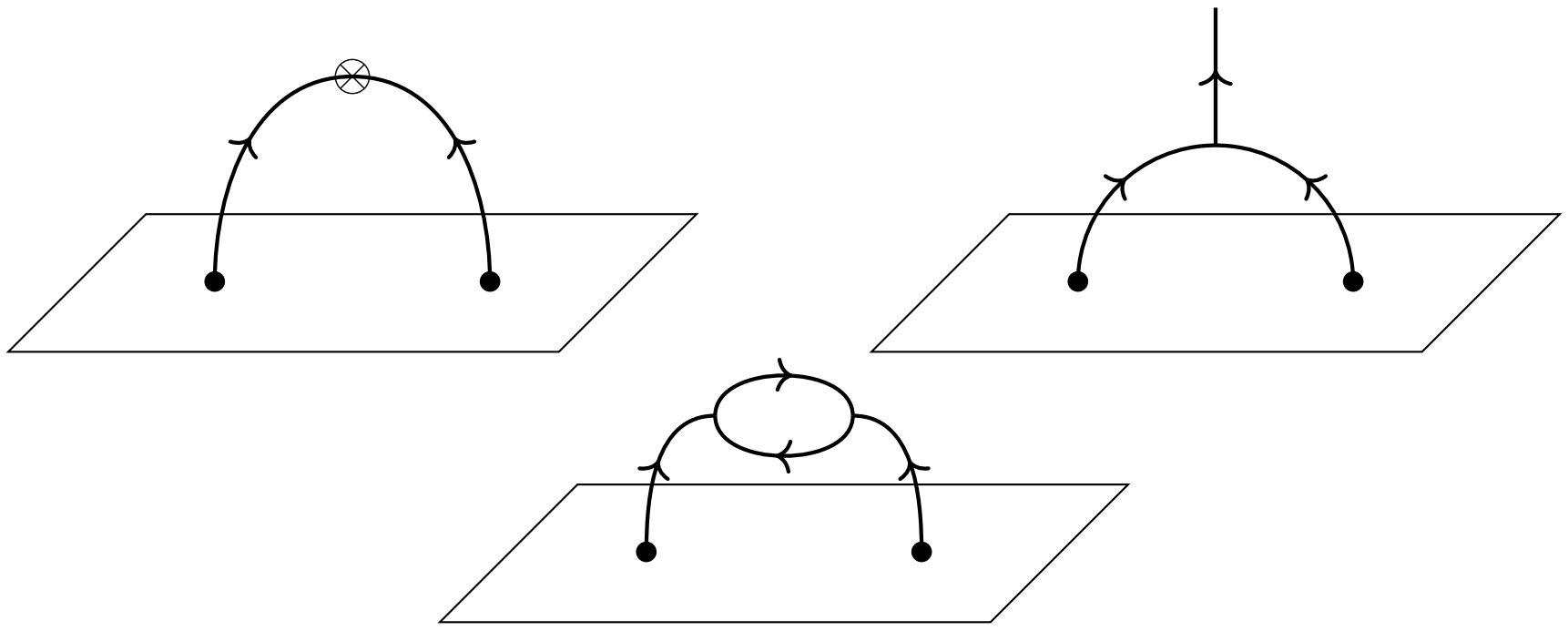
\begin{figure}[ht]
		\centering
		\begin{tikzpicture}
			\node (int) at (0,1.5) {$\otimes$};
			\node (b1) at (-1,0) {$\bullet$};
			\node (b2) at (1,0) {$\bullet$};
			
			\draw[fermion, thick] (b1.center) to [in=180, out=90] (int.center);
			\draw[fermion, thick] (b2.center) to [in=0, out=90] (int.center);
			\draw (-2.5,-0.5) -- (-1.5,0.5) -- (2.5,0.5) -- (1.5,-0.5) -- cycle;
		\end{tikzpicture} \hspace{1cm} \begin{tikzpicture}
			\node (int) at (0,1) {};
			\node (b1) at (-1,-0) {$\bullet$};
			\node (b2) at (1,0) {$\bullet$};
			\node (b3) at (0,2) {};
			
			\draw[fermion, thick] (b1.center) to [in=180, out=90] (int.center);
			\draw[fermion, thick] (b2.center) to [in=0, out=90] (int.center);
			\draw[fermion, thick] (int.center) to [in=270, out=90] (b3.center);
			\draw (-2.5,-0.5) -- (-1.5,0.5) -- (2.5,0.5) -- (1.5,-0.5) -- cycle;
		\end{tikzpicture} \hspace{1cm} \begin{tikzpicture}
			\node (int1) at (-0.5,1) {};
			\node (int2) at (0.5,1) {};
			\node (b1) at (-1,-0) {$\bullet$};
			\node (b2) at (1,0) {$\bullet$};
			
			\draw[fermion, thick] (b1.center) to [in=180, out=90] (int1.center);
			\draw[fermion, thick] (b2.center) to [in=0, out=90] (int2.center);
			\draw[fermion, thick] (int1.center) to [in=90, out=90] (int2.center);
			\draw[fermion, thick] (int2.center) to [in=270, out=270] (int1.center);
			\draw (-2.5,-0.5) -- (-1.5,0.5) -- (2.5,0.5) -- (1.5,-0.5) -- cycle;
		\end{tikzpicture}
		\caption{An illustration of the three Feynman diagrams contributing to the singular part of the OPE of $\bB$ and itself.}
		\label{fig:affineFeynman}
	\end{figure}
	
	If we put these computations together we find that the OPE takes the form
	\be
	\begin{aligned}
		B_a(z) B_b(w) & \sim \hbar \bigg(\frac{1}{(z-w)^2}\frac{-k K_{ab}}{4\pi^2} + \frac{1}{z-w} \frac{f^c_{ab} B_c(w)}{2\pi \ii}\bigg)\\
		& \qquad + \hbar^2\bigg(\frac{1}{(z-w)^2}\frac{h^\vee K_{ab}}{4\pi^2}\bigg)
	\end{aligned}
	\ee
	where $B_a$ is the zero-form part of $\bB_a|$.
	If we set $J_a = (2\pi \ii) B_a$ then we see that the OPEs of the $J$'s furnish an affine $\fg$ current algebra at level $k_\hbar = k - h^\vee \hbar$, i.e. the $\hbar$-adic vertex algebra $V_{k_\hbar}(\fg)$ freely generated by the $J$'s.
	In particular, we see that the 1-loop correction shifts the classical level as $k \to k - h^\vee \hbar$, as claimed.
\end{proof}

The shifted double pole is also obtained by replacing $k$ by $k_\hbar$ in the classical quadratic vertex.
The all-Dirichlet specialization of the effective-action calculation in Appendix~\ref{app:EffectiveAction} gives the corresponding local chiral correction:
\be
    -\frac{\ii\hbar h^\vee}{4\pi}\int\d z~\Tr(\bA\partial_z\bA).
\ee
Throughout the paper, $k$ denotes the classical coupling and $k_\hbar$ the affine level obtained from the chosen quantization.

\section{Affine \texorpdfstring{$W$}{W}-algebras from bulk-boundary quantization}\label{sec:Walgebras}

We now quantize the principal classical affine $W$-algebra associated to a simple complex Lie algebra $\fg$.
The key step is to use a quasi-isomorphic dg Poisson vertex algebra with linear brackets.
Its universal bulk is Chern--Simons theory expanded about a principal-nilpotent background, with an oper boundary condition.
The holomorphic-gauge quantization gives a dg vertex algebra with explicitly computed OPEs and differential.
Its cohomology is $W_{k-h^\vee\hbar}(\fg)$, as stated in Theorem~\ref{thm:Walgebra}.
This is the explicit presentation of the deformation quantization used below.

Section~\ref{sec:ClassicalWalgebra} constructs the reduced Drinfel'd--Sokolov complex and identifies its universal bulk.
Section~\ref{sec:QuantumWalgebra} computes the quantum boundary complex and compares it with the standard quantum Drinfel'd--Sokolov complex.

\subsection{The classical affine $W$-algebra and its universal bulk theory}\label{sec:ClassicalWalgebra}

The classical affine $W$-algebra of $\fg$ at level $k$ is most naturally defined cohomologically in terms of classical Drinfel'd--Sokolov reduction \cite{Drinfeld:1985ds} of the affine Poisson vertex algebra $\CV_k(\fg)$ which we now review.
We largely follow the notation of \cite{FBZ}, cf. Chapter 15 thereof.

We first choose a triangular decomposition $\fg = \fn_- \oplus \mathfrak{h} \oplus \fn_+$.
Let $\Delta_+$ denote the set of positive roots with respect to $\mathfrak{h}$; for a root $\alpha \in \Delta_+$ let $e_\alpha \in \fn_+$ denote the corresponding root vector.
We choose a basis $\{T_{\ol{a}}\}$ of $\mathfrak{b}_- = \fn_- \oplus \mathfrak{h}$ that consists of the Cartan elements $h_i \in \mathfrak{h}$ and the root vectors $f_\alpha \in \fn_-$; we define a basis $\{T_a\}$ of $\fg$ as the union $\{e_\alpha\} \cup \{T_{\ol a}\}$.
We denote by $f^a_{b c}$ the structure constants of $\fg$ in this basis:
\be
	[T_a, T_b] = f^c_{a b} T_c~.
\ee
Finally, define the character $\chi: \fn_+ \to \C$ such that
\be
	\chi(e_\alpha) = \begin{cases}
		1 & \mathrm{if}~\alpha~\mathrm{is~simple}\\
		0 & \mathrm{else}
	\end{cases}~.
\ee

In addition to the affine Poisson vertex algebra we consider a Clifford Poisson vertex algebra valued in $\fn_+$:
\be
	\wt{\CC}_k(\fg) = \CV_k(\fg) \otimes \CC\ell(\fn_+)~.
\ee
Explicitly, we introduce Grassmann odd fields $\psi^\alpha$, $\ol{\psi}_\alpha$ with $\lambda$-bracket
\be
	\{\psi^\alpha {}_\lambda \ol{\psi}_\beta\} = \delta^\alpha{}_\beta~.
\ee
We will also denote the fields generating $\CV_k(\fg)$ by $\wt{J}_a$ instead of $J_a$.
We then form the field
\be
	\mathcal{Q} = \psi^\alpha (\wt{J}_\alpha - \chi(e_\alpha)) - \tfrac{1}{2} f^{\gamma}_{\alpha \beta} \ol{\psi}_\gamma \psi^\alpha \psi^\beta
\ee
and use it to define the differential
\be
	Q_{\mathrm{cl}} = \{\mathcal{Q} {}_0 -\}~.
\ee
If we give $\wt{J}_a$ degree $0$, $\psi^\alpha$ degree $1$, and $\ol{\psi}_\alpha$ degree $-1$ then $Q_{\mathrm{cl}}$ equips $\wt{\CC}_k(\fg)$ with the structure of a dg Poisson vertex algebra.
The classical $W$-algebra of type $\fg$ is precisely the zeroth cohomology of $\wt{\CC}_k(\fg)$:
\be
	\CW_k(\fg) := H^0(\wt{\CC}_k(\fg), Q_{\mathrm{cl}})~.
\ee
Just as with the quantum $W$-algebras, the cohomology of this dg Poisson vertex algebra is concentrated in degree $0$.
For the standard classical Drinfel'd--Sokolov computation in unit-level normalization, see \cite[Sec.~3.5, Example~3.20]{DeSole:2012pva}.

We define
\be
	J_a = \wt{J}_a + f^\beta_{a \alpha} \psi^\alpha \ol{\psi}_\beta~.
\ee
It is clear that the $J_a$, $\psi^\alpha$, $\ol{\psi}_\alpha$ also generate the underlying Poisson vertex algebra.
Most importantly, the subalgebra generated by $\wh{J}_\alpha$ and $\ol{\psi}_\alpha$ forms an acyclic subcomplex.
After removing it, we are left with a smaller, quasi-isomorphic dg Poisson vertex algebra $\CC_k(\fg)$ generated by $J_{\ol{a}}$ and $\psi^\alpha$; see e.g. \cite{dBT94, FBZ} for the quantum version of this statement.
The $\lambda$-bracket of these fields takes the form
\be\label{eq:DSBrackets}
	\{J_{\ol a} {}_\lambda J_{\ol b}\} = f^{\ol c}_{\ol{a} \ol{b}} J_{\ol{c}} + \lambda k K_{\ol{a} \ol{b}} \qquad \{J_{\ol a} {}_\lambda \psi^\alpha\} = -f^{\alpha}_{\ol{a} \beta} \psi^\beta \qquad \{\psi^\alpha {}_\lambda \psi^\beta\} = 0
\ee
and differential is given by
\be\label{eq:DSDifferential}
	Q_{\mathrm{cl}} J_{\ol a} = f^{\ol b}_{\alpha \ol{a}} \psi^\alpha J_{\ol b} + f^{\alpha}_{\beta \ol{a}} \chi(e_\alpha) \psi^\beta + k K_{\ol{a} \alpha} \partial \psi^\alpha \qquad Q_{\mathrm{cl}} \psi^\alpha = -\tfrac{1}{2} f^\alpha_{\beta \gamma} \psi^\beta \psi^\gamma~.
\ee

With the above dg model for the classical $W$-algebra in hand, we describe its universal bulk theory.
To construct the universal bulk, extend $J_{\ol a}$ and $\psi^\alpha$ to fields
\be
	\bJ_{\ol{a}} \d z^{\otimes 1-\mathrm{pg}(T_{\ol{a}})} \in \bOmega^{(1-\mathrm{pg}(T_{\ol{a}})), \bullet} \qquad \boldsymbol{\psi}^\alpha \d z^{\otimes\mathrm{pg}(e_\alpha)} \in \bOmega^{(\mathrm{pg}(e_\alpha)), \bullet}[1]~,
\ee
whose restriction to the boundary give rise to the above $J_{\ol a}$ and $\psi^\alpha$, where $\mathrm{pg}(A)$ for $A \in \fg$ is the principal gradation defined by
\be
	\mathrm{pg}(h_i) = 0 \qquad \mathrm{pg}(e_\alpha) = \mathrm{height}(\alpha) \qquad \mathrm{pg}(f_\alpha) = - \mathrm{height}(\alpha)~.
\ee
We also introduce their conjugate fields
\be
	\boldsymbol{\beta}^{\ol{a}}_J \partial_z^{\otimes-\mathrm{pg}(T_{\ol a})} \in \bOmega^{(\mathrm{pg}(T_{\ol a})), \bullet}[1] \qquad \boldsymbol{\beta}_{\psi, \alpha} \partial_z^{\otimes \mathrm{pg}(e_\alpha)-1} \in \bOmega^{(1-\mathrm{pg}(e_\alpha)), \bullet}~.
\ee

The canonical boundary condition sets all of the $\boldsymbol{\beta}$ fields to zero.
The construction of \cite[Secs.~2.2 and 4.1]{Khan:2025rah} encodes the $\lambda$-bracket in the bulk action;
for a dg Poisson vertex algebra one additionally includes the cotangent lift of its differential.
In the present example this gives
\be\label{eq:UniversalBulkAction}
\begin{aligned}
	S_{\mathrm{univ}} & = \int \d z \bigg\{\boldsymbol{\beta}^{\ol a}_J \d' \bJ_{\ol{a}} + \boldsymbol{\beta}_{\psi,\alpha} \d' \boldsymbol{\psi}^\alpha\\
	& \qquad +2\pi \ii \left(\tfrac{1}{2}  \boldsymbol{\beta}^{\ol a}_J \boldsymbol{\beta}^{\ol b}_J(f^{\ol c}_{\ol{a} \ol{b}} \bJ_{\ol{c}}) - \tfrac{1}{2} \boldsymbol{\beta}^{\ol a} \partial_z \boldsymbol{\beta}^{\ol b}(k K_{\ol{a} \ol{b}}) + \boldsymbol{\beta}^{\ol a}_J \boldsymbol{\beta}_{\psi, \alpha}(f^{\alpha}_{\ol{a} \beta} \boldsymbol{\psi}^\beta)\right)\\
	& \qquad + \boldsymbol{\beta}^{\ol a}_J(f^{\ol b}_{\alpha \ol{a}} \boldsymbol{\psi}^\alpha \bJ_{\ol b} + f^{\alpha}_{\beta \ol{a}} \chi(e_\alpha) \boldsymbol{\psi}^\beta + k K_{\ol{a} \alpha} \partial_z \boldsymbol{\psi}^\alpha) + \boldsymbol{\beta}_{\psi, \alpha} (-\tfrac{1}{2} f^{\alpha}_{\beta \gamma} \boldsymbol{\psi}^\beta \boldsymbol{\psi}^\gamma)\bigg\}
\end{aligned}~.
\ee
The first line is the kinetic term, the second encodes the brackets in Eq. \eqref{eq:DSBrackets}, and the third is the Hamiltonian $S_{Q_{\mathrm{cl}}}$ for the cotangent lift of the differential in Eq. \eqref{eq:DSDifferential}.
In particular, $\{S_{Q_{\mathrm{cl}}},-\}_{\mathrm{BV}}$ acts on $\bJ_{\ol a},\boldsymbol{\psi}^\alpha$ by the displayed differential and on their partners by its cotangent action.
The identities $Q_{\mathrm{cl}}^2=0$ and compatibility of $Q_{\mathrm{cl}}$ with the $\lambda$-bracket say respectively that $\{S_{Q_{\mathrm{cl}}},S_{Q_{\mathrm{cl}}}\}_{\mathrm{BV}}=0$ and that $S_{Q_{\mathrm{cl}}}$ is compatible with the Poisson interaction in the second line.
Thus this term incorporates the BRST reduction into the three-dimensional BV theory.

With some algebra, this action can be rearranged to be precisely the mixed BF realization of Chern--Simons theory at level $k$ expanded about a non-trivial, constant value for $\bB_{\alpha}$ proportional to $\chi(e_\alpha)$.
Correspondingly, we will refer to this field theory as Higgsed Chern--Simons theory.
Explicitly, we identify the fields as
\be\label{eq:UniversalCSIdentification}
\begin{aligned}
	\bA^{\ol a} & = 2\pi \ii~\boldsymbol{\beta}^{\ol a}_J & \qquad \bA^\alpha & = - \boldsymbol{\psi}^\alpha\\
	\bB_{\ol a} & = (2\pi \ii)^{-1} \bJ_{\ol a} & \qquad \bB_\alpha & = - \boldsymbol{\beta}_{\psi, \alpha} + \frac{1}{2\pi\ii} \chi(e_\alpha)\\
\end{aligned}~.
\ee
Substituting Eq. \eqref{eq:UniversalCSIdentification} into the Chern--Simons action reproduces Eq. \eqref{eq:UniversalBulkAction} after integration by parts and the boundary term implementing this choice of polarization.
In these variables the boundary condition fixes
\be
	\bA^{\ol a}| = 0 \qquad \bB_{\alpha}| = \frac{1}{2 \pi \ii}\chi(e_\alpha)~.
\ee
We refer to this as an \enquote{oper} boundary condition for perturbative Chern--Simons theory.
Indeed, identifying $\bB_a$ with the $\d z$ part of a $\fg$-connection we see that this boundary condition enforces that the boundary value of this connection is an oper.

We can now identify precisely the three-dimensional term responsible for the Higgsing.
The shift $\bB_\alpha=b_\alpha-\boldsymbol{\beta}_{\psi, \alpha}$, where $b_\alpha = (2\pi \ii)^{-1} \chi(e_\alpha)$, in the Chern--Simons action adds the background-dependent term
\be
	\Delta S_b=b_\alpha\int\d z~\big(\d'\bA^\alpha + \tfrac{1}{2} f^{\alpha}_{\beta \gamma} \bA^\beta \bA^\gamma + f^\alpha_{\ol{a} \beta}\bA^{\ol a}\bA^\beta\big).
\ee
The first summand is a total derivative; on a manifold with boundary it is canceled by the boundary term
$-b_\alpha\int_{\partial M}\bA^\alpha$ appropriate to fixing $\bB_\alpha|=b_\alpha$, with the induced boundary orientation.
The second summand vanishes because $\chi$ is a character for $\fn_+$.
The resulting bulk deformation is
\be\label{eq:HiggsBulkTerm}
	S_{\mathrm{Higgs}}= b_\alpha f^\alpha_{\ol{a} \beta} \int\d z~ \bA^{\ol a}\bA^\beta	= \chi(e_\alpha) f^\alpha_{\beta \ol{a}} \int\d z~\boldsymbol{\beta}^{\ol a}_J\boldsymbol{\psi}^\beta.
\ee
The other two summands of $S_Q$ already occur in the unshifted Chern--Simons action: $f^{\ol{b}}_{\alpha \ol{a}}\boldsymbol{\beta}^{\ol a}_J \bJ_{\ol b} \boldsymbol{\psi}^\alpha$ comes from $f^{\ol{b}}_{\alpha \ol{a}} \bB_{\ol{a}} \bA^\alpha \bA^{\ol{a}}$, while $k K_{\ol{a} \alpha} \boldsymbol{\beta}^{\ol a}_J\partial_z \boldsymbol{\psi}^\alpha$ comes from the mixed holomorphic Chern--Simons term.
The quadratic deformation in Eq. \eqref{eq:HiggsBulkTerm} introduces the algebraic differential pairing
$\bJ_{\ol a}$ with $\boldsymbol{\psi}^\alpha$, and also pairs their cotangent partners.

Geometrically, the trace pairing identifies $\chi(e_\alpha) \in \fg^\vee$ with a sum over $f_\alpha$ for simple roots $\alpha$, so the vacuum $\bB_\alpha = b_\alpha$ is a regular nilpotent background.
It is the principal-nilpotent background entering the oper description of the boundary condition \cite[Sec.~4.4]{Butson:2016jys}; see also \cite[Sec.~3.2]{Khan:2025rah} for $\fg = \fsl_2$.

\subsection{Quantized oper boundary conditions and affine $W$-algebras}
\label{sec:QuantumWalgebra}

Having identified the universal bulk theory for the classical affine $W$-algebra $\CW_k(\fg)$ with Chern--Simons theory expanded around a certain field configuration, we now proceed with perturbative quantization thereof.
Namely, we start with the above oper boundary condition of perturbative Chern--Simons theory and then quantize the algebra of boundary observables.

The holomorphic-gauge construction and anomaly-vanishing argument of \cite[Secs.~5 and 6]{Gwilliam:2019cbp} extend to the oper boundary condition, giving a finite, 1-loop exact solution to the quantum master equation.
The mixed-boundary adaptation uses the vanishing of the two-vertex color trace when both inputs lie in $\fn_+$, since their adjoint actions strictly raise principal grading.

With this quantization in hand, we compute the local chiral quadratic vertex needed for the quantum boundary differential.
We then determine the boundary OPEs and identify the resulting dg vertex algebra with the reduced Drinfel'd--Sokolov complex.
Here \enquote{local chiral} refers to the Taylor coefficient retaining holomorphic derivatives of the external fields, with no antiholomorphic or normal derivatives, using the uncut propagators of Appendix~\ref{app:EffectiveAction}.
It does not mean that the full effective interaction is local or quadratic.

\begin{thm}\label{thm:oper1loop}
	The local chiral quadratic part of the 1-loop effective interaction in holomorphic gauge is
	\be
		\frac{\mathrm{i} \hbar}{4\pi}  \int \d z ~ B(\bA, \partial_z \bA)
	\ee
	where the symmetric bilinear form $B \in \Sym^2(\fg^\vee)$ is given by
	\be
	\begin{aligned}
		B(h_i, h_j) & = 0 & \qquad B(h_i, e_{\alpha}) & = 0 = B(h_i, f_{\alpha})\\
		B(e_\alpha, f_\beta) & = -\delta_{\alpha \beta} & \qquad B(e_\alpha, e_\beta) & = 0 = B(f_\alpha, f_\beta)
	\end{aligned}
	\ee
	for all $i,j \in \{1, \dots, \ell\}$ and $\alpha,\beta \in \Delta_+$, with root vectors normalized so that $[e_\alpha,f_\alpha]=h_\alpha$ and $\alpha(h_\alpha)=2$.
\end{thm}

\begin{proof}
	By the graph count of \cite[Sec.~1.2]{Gwilliam:2019cbp}, the 1-loop interaction is independent of $\bB$.
	For its local chiral part, each $\bA$ has total degree one and integration lowers degree by two, so a degree-zero functional is quadratic in $\bA$.
	The corresponding graph is the two-vertex wheel,\footnote{The term \emph{wheel} is used inconsistently in the literature. In the present work, like in \cite{Gwilliam:2019cbp} a $n$-point wheel is a one-loop graph on $n$-vertices, potentially including external legs. The version without legs is usually called \emph{cycle} in the graph theory literature. Conversely, much of the literature on  Feynman integral computations, e.g. \cite{gurdogan_new_2016}, uses the term \emph{wheel with $n$ spokes} for a class of graphs with higher loop order, consisting of one central vertex, $n$ outgoing \enquote{spokes}, and a number of concentric cycles.  }
	Angular integration retains precisely the first holomorphic Taylor coefficient, giving a term $\int\d z~B(\bA,\partial_z\bA)$.
	This is the first-moment mechanism of \cite[Lemma~6.3]{Gwilliam:2019cbp}; Appendix~\ref{app:EffectiveAction} implements it for the mixed boundary kernels.
	This argument does not exclude nonlocal higher-point wheels.
	
	We are thus left with showing that this bilinear form is given as in Theorem \ref{thm:oper1loop}.
	There are two parts to this: there is a purely Lie-algebraic factor as well as a factor coming from the underlying Feynman integral.
	We compute the latter in Appendix \ref{app:EffectiveAction}.
	The result of these integrals implies that the bilinear form is given by
	\be
	B(T_a, T_b) = \tfrac{1}{2} \sum_{\alpha, \beta} f^\alpha_{a\beta} f^\beta_{b\alpha} - \tfrac{1}{2} \sum_{\ol{a}, \ol{b}} f^{\ol a}_{a \ol{b}} f^{\ol b}_{b \ol{a}}~.
	\ee
	If we denote by $P_+$ the projection $\fg \to \fn_+$ and denote $P_- := \mathrm{Id}_\fg - P_+$ the projection onto $\fn_- \oplus \mathfrak{h}$ then we can express the above as
	\be
	B(T_a, T_b) = \tfrac{1}{2}\Tr_{\fg}(P_+ \mathrm{ad}_{T_a} P_+ \mathrm{ad}_{T_b}) - \tfrac{1}{2}\Tr_{\fg}(P_- \mathrm{ad}_{T_a} P_- \mathrm{ad}_{T_b})~.
	\ee
	As the projections $P_\pm$ commute with the adjoint action of $\mathfrak{h}$, we see that this expression is $\mathfrak{h}$-invariant.
	Moreover, if we take $T_a, T_b \in \mathfrak{h}$ then the two terms cancel each other as the nonzero weights of $\mathrm{ad}_{T_a}$ on $\fn_+$ are the negative of those on $\fn_-$ and so $B$ vanishes upon restriction to $\mathfrak{h}$.
	
	To finish, let $P_0$ and $P_{\fn_-}$ denote the projections onto $\mathfrak{h}$ and $\fn_-$, respectively.
	Transpose with respect to the invariant form gives $P_+^t=P_{\fn_-}$ and $(\mathrm{ad}_X)^t=-\mathrm{ad}_X$.
	Thus the traces supported entirely on $\fn_+$ and $\fn_-$ cancel.
	Expanding $P_-=P_{\fn_-}+P_0$ and using $P_0\mathrm{ad}_X P_0=0$, we obtain
	\be
	 B(X,Y)=-\tfrac12\Tr_{\mathfrak h}\big(P_0\mathrm{ad}_X P_{\fn_-}\mathrm{ad}_Y
	                  +P_0\mathrm{ad}_Y P_{\fn_-}\mathrm{ad}_X\big)~.
	\ee
	For $X=e_\alpha$ and $Y=f_\alpha$, only the first summand survives; it sends $h\in\mathfrak h$ to $\alpha(h)[e_\alpha,f_\alpha]$.
	Its trace is $\alpha(h_\alpha)=2$, so $B(e_\alpha,f_\alpha)=-1$ as claimed.
\end{proof}

\begin{rmk}\label{rmk:OperPolarization}
The oper boundary condition is not preserved by the full adjoint action of $\fg$, so the propagators and the resulting bilinear form $B$ need not be $\fg$-invariant.
The QME applies to the complete effective interaction and does not require its extracted local quadratic part to be separately BV-closed.
\end{rmk}

We now compute the vertex algebra of boundary observables.
For the oper boundary condition, the elementary OPEs retain their tree-level form; the level shift will enter through the quantum differential and normal ordering.
Counting $\bB$-inputs excludes loop corrections to OPEs involving an $A^\alpha$ observable, while ghost degree forces a loop correction to the $B_{\ol a}B_{\ol b}$ OPE to be central.
The two-point calculation gives no such correction, since it is proportional to $B|_{\mathfrak b_-\otimes\mathfrak b_-}=0$.
Using the Feynman integrals computed in Appendix \ref{app:AffineOPEs}, we find that the singular terms in the OPEs are given by
\be
\begin{aligned}
	B_{\ol{a}}(z) B_{\ol{b}}(w) & \sim \hbar\left(\frac{1}{(z-w)^2}\frac{-k K_{\ol{a} \ol{b}}}{4 \pi^2} + \frac{1}{z-w} \frac{f^{\ol{c}}_{\ol{a}\ol{b}}B_{\ol{c}}}{2\pi\ii}\right)\\
	B_{\ol{a}}(z) A^{\alpha}(w) & \sim \hbar \left(\frac{1}{z-w}\frac{-f^{\alpha}_{\ol{a} \beta} A^\beta(w)}{2 \pi \ii}\right)\\
	A^{\alpha}(z) A^{\beta}(w) & \sim 0
\end{aligned}~.
\ee
The 1-loop correction we found above appears in the differential.
As the OPE of $A$'s with themselves is regular, the differential's action on $A^\alpha$ takes the same form as in the classical observables:
\be
	Q_{\hbar} A^\alpha(z) = \tfrac{1}{2} f^\alpha_{\beta \gamma} :A^\beta A^\gamma:(z)
\ee
The differential's action on $B_{\ol a}$ differs from the classical expression in two ways: we must account for the local quadratic vertex computed above and normally order the quadratic term.
It is thus given by
\be
\begin{aligned}
	Q_{\hbar} B_{\ol a}(z) & = \tfrac{1}{2}f^{\ol b}_{\ol{a} \alpha} \big(:A^\alpha B_{\ol b}:(z) + :B_{\ol b} A^\alpha:(z)\big)\\
	& \qquad - \frac{\ii}{2\pi}f^{\alpha}_{\ol{a} \beta} \chi(e_\alpha) A^\beta(z) + \frac{\ii}{2\pi}\big(k K_{\ol{a} \alpha}+\hbar B(T_{\ol a},e_\alpha)\big) \partial_z A^\alpha(z)~.
\end{aligned}
\ee

In order to identify the above OPEs with a deformation quantization of the classical affine $W$-algebra, we set
\be
	J_{\ol a} = (2\pi \ii) B_{\ol a} \qquad \psi^\alpha = - A^\alpha~.
\ee
With this choice, we find that the OPEs are given by
\be
\begin{aligned}
	J_{\ol{a}}(z) J_{\ol{b}}(w) & \sim \hbar\left(\frac{k K_{\ol{a} \ol{b}}}{(z-w)^2} + \frac{f^{\ol{c}}_{\ol{a}\ol{b}}J_{\ol{c}}}{z-w}\right)\\
	J_{\ol{a}}(z) \psi^{\alpha}(w) & \sim \hbar \left(\frac{-f^{\alpha}_{\ol{a} \beta} \psi^\beta(w)}{z-w}\right)\\
	\psi^{\alpha}(z) \psi^{\beta}(w) & \sim 0
\end{aligned}
\ee
which reproduce the $\lambda$-brackets given in Eq. \eqref{eq:DSBrackets} in the classical limit.
The differential takes the form
\be
\begin{aligned}
	Q_{\hbar} J_{\ol a}(z) & = \tfrac{1}{2}f^{\ol b}_{\alpha \ol{a}} \big(:\psi^\alpha J_{\ol b}:(z) + :J_{\ol b} \psi^\alpha:(z)\big)\\
	& \qquad + f^{\alpha}_{\beta \ol{a}} \chi(e_\alpha) \psi^\beta(z) + \big(k K_{\ol{a} \alpha}+\hbar B(T_{\ol a},e_\alpha)\big) \partial_z \psi^\alpha(z)\\
	Q_{\hbar} \psi^\alpha(z) & = -\tfrac{1}{2} f^{\alpha}_{\beta \gamma} :\psi^\beta \psi^\gamma:
\end{aligned}
\ee
which reproduces the differential $Q_{\mathrm{cl}}$ given in Eq. \eqref{eq:DSDifferential} in the classical limit.
Keeping $B$ explicit accounts for root lengths: in the above normalization,
$B(f_\alpha,e_\alpha)=-1$, whereas $K(f_\alpha,e_\alpha)=2/(\alpha,\alpha)$.
Thus the coefficient simplifies to $(k-\hbar)K_{\ol a\alpha}$ when $\fg$ is simply laced, with long roots of squared length two.
To compare with the usual Drinfel'd--Sokolov differential, put
\be
 M(X,Y)=\Tr_{\fg}(P_+\mathrm{ad}_X P_+\mathrm{ad}_Y),\qquad
 C_{\ol a\beta}=\sum_{\alpha,\ol b}f^{\ol b}_{\alpha\ol a}f^\alpha_{\ol b\beta}~.
\ee
Reordering the symmetrized quadratic term gives $f^{\ol b}_{\alpha\ol a}:\psi^\alpha J_{\ol b}:$ and subtracts $\hbar C_{\ol a\beta}\partial_z\psi^\beta/2$.
Since $\mathrm{ad}_{T_{\ol a}}$ preserves $\mathfrak b_-$, decomposition of the Killing trace gives
$B(T_{\ol a},e_\beta)-C_{\ol a\beta}/2=M(T_{\ol a},e_\beta)-h^\vee K_{\ol a\beta}$.
The derivative coefficient is therefore $(k-\hbar h^\vee)K_{\ol a\beta}+\hbar M(T_{\ol a},e_\beta)$, as in the reduced Drinfel'd--Sokolov differential of \cite[Eq.~(5.14)]{DeSole:2006finite}.
Also $M|_{\mathfrak b_-\otimes\mathfrak b_-}=h^\vee K|_{\mathfrak b_-\otimes\mathfrak b_-}$, so the displayed OPEs agree with Eq.~(5.10) of loc.\ cit.
Thus the elementary OPEs and the corrected differential together give the reduced Drinfel'd--Sokolov complex at affine level $k-h^\vee\hbar$.
The level shift found for Dirichlet boundary conditions is therefore recovered in the oper presentation through the differential.
Taking cohomology proves the following analogue of Theorem~\ref{thm:1loopaffine}.

\begin{thm}\label{thm:Walgebra}
	The holomorphic gauge-fixing of perturbative Chern--Simons theory on $\C\times\R_{\geq0}$ at classical level $k$ with oper boundary conditions determines a vertex algebra of quantum boundary observables isomorphic to the $\hbar$-adic principal affine $W$-algebra
	\be
		W_{k_\hbar}(\fg)~.
	\ee
	The 1-loop correction shifts the affine level entering Drinfel'd--Sokolov reduction from $k$ to $k_\hbar = k - h^\vee \hbar $, where $h^\vee$ is the dual Coxeter number of $\fg$.
\end{thm}

\section{\texorpdfstring{$\mathfrak{sl}_2$}{sl2} Chern--Simons theory and chiral gravity}\label{sec:CSgravity}

The Virasoro Poisson vertex algebra has two bulk presentations in our framework: three-dimensional chiral gravity and $\fsl_2$ Chern--Simons theory with the oper boundary condition.
We compare them first classically and then quantum mechanically, obtaining a chiral counterpart of the Chern--Simons formulation of three-dimensional gravity \cite{Witten:1988hc}.
At the quantum level, the two constructions yield Virasoro algebras with different expressions for the central charge; matching them identifies the corresponding choices of quantization.

\subsection{Three-dimensional chiral gravity}\label{sec:ClassicalGravity}

The first field theory we will be interested in is a chiral version of three-dimensional gravity.
Formulated as a BV field theory, this theory has two fields
\be
	\bmu \partial_z \in \bOmega^{(-1),\bullet}[1] \qquad \bG\, \d z^{\otimes 2}\in \bOmega^{(2),\bullet}~
\ee
with action functional
\be
	S = \int\d z \left\{\bG \big(\d' \bmu + \bmu \partial_z \bmu\big) + \frac{\ii c}{48 \pi} \bmu \partial_z^3 \bmu\right\}
\ee
and BV symplectic form
\be
	\Omega_{BV} = \int\d z~\delta \bG \delta \bmu~.
\ee
The resulting equations of motion are encoded in the BV differential
\be
	Q \bmu = \d' \bmu + \bmu \partial_z \bmu \qquad Q \bG = \d' \bG + 2 \bG \partial_z \bmu + \bmu \partial_z \bG + \frac{\ii c}{24 \pi} \partial_z^3 \bmu~.
\ee

We note that this field theory can be realized as the universal bulk theory associated to the Virasoro Poisson vertex algebra, cf. Section 3.2 of \cite{Khan:2025rah}.
This Poisson vertex algebra is freely generated by a single field $T$ with $\lambda$-bracket given by
\be
	\{T{}_\lambda T\} = \partial T + \lambda (2 T) + \lambda^3 \bigg(\frac{c}{12}\bigg)~.
\ee
Dually, the Virasoro Poisson vertex algebra arises as the classical observables on a Dirichlet boundary condition defined by $\bmu| = 0$; as we will see in Section \ref{sec:Virasoro} the generator $T$ is realized by the boundary value of the field $\bG$, up to an overall numerical factor.

There are two reasons why we refer to this theory as a gravitational theory.
First and foremost, this theory can be viewed as studying deformations of the transverse holomorphic foliation of the underlying spacetime: the equation of motion $\d' \bmu + \bmu \partial_z \bmu = 0$ implies that the combination $\d' + \bmu \partial_z$ defines a new, integrable transverse holomorphic foliation.
In other words, the field $\bmu$ is the transverse holomorphic foliation analogue of a Beltrami differential.
We expect this theory to arise from the holomorphic--topological twist of three-dimensional $\mathcal N=2$ conformal supergravity; we discuss the evidence and the remaining questions in Section~\ref{sec:Outlook}.

The second connection to three-dimensional gravitational theories can be seen by inspecting the moduli space of solutions to the equations of motion, cf. \cite[Sec.~3.2]{Khan:2025rah}.
If we place the theory on a three-manifold of the form $\R \times \Sigma_g$, where $\Sigma_g$ is a Riemann surface of genus $g$, then this moduli space of solutions can be identified with a $c$-dependent twist of the cotangent bundle to Teichm\"{u}ller space of complex structures; the field $\bmu$ can be viewed as a Beltrami differential on $\Sigma_g$.%
\footnote{One could also consider a theory that leads to a twisted cotangent bundle to the moduli space of genus $g$ Riemann surfaces, instead of Teichm\"{u}ller space, by quotienting the action of the mapping class group on $\Sigma_g$. This can be viewed as a choice of whether to quotient by large gauge transformations or, rather, the appropriate space of fields one integrates over.} %
Geometric quantization of the twisted cotangent bundle Teichm\"{u}ller space is known to lead to Virasoro conformal blocks \cite{Verlinde:1989ua, Teschner:2003em}; as noted in \cite{Khan:2025rah}, the 1-loop correction to the level we identify below can be viewed as the metaplectic correction necessary for geometric quantization.
These conformal blocks, in turn, are an essential ingredient in constructing the Virasoro TQFT of \cite{Collier:2023fwi, Collier:2024mgv} which can be viewed as a half of pure three-dimensional gravity with negative cosmological constant.
Thus three-dimensional chiral gravity is a natural candidate for a Lagrangian presentation of Virasoro TQFT.

\subsection{Classical boundary identification}\label{sec:ClassicalIdentification}

Of course, the Virasoro Poisson vertex algebra is also the classical affine $W$-algebra for $\fsl_2$.
Making explicit the dg Poisson vertex algebra described in Section \ref{sec:ClassicalWalgebra} in this case, we start with a Poisson vertex algebra freely generated by two degree 0, bosonic fields $h, f$ and a degree 1, fermionic field $\psi$ and equipped with the differential
\be
	Q_{\mathrm{cl}} h = -2 \psi \qquad Q_{\mathrm{cl}} f = h \psi + k \partial \psi \qquad Q_{\mathrm{cl}} \psi = 0
\ee
extended by the Leibniz rule; the nonzero $\lambda$-brackets are given by
\be
	\{h {}_\lambda h\} = \lambda (2 k) \qquad \{h {}_\lambda f\} = -2 f \qquad \{h {}_\lambda \psi\} = -2\psi~.
\ee
For $k \neq 0$, the cohomology of $Q_{\mathrm{cl}}$ is generated by the field
\be
	T = k^{-1} \left(f +  \tfrac{1}{4} h^2 + \tfrac{1}{2} k \partial h\right)
\ee
whose $\lambda$-bracket with itself is given by
\be
	\{T{}_\lambda T\} = \partial T + \lambda (2 T) + \lambda^3\bigg(\frac{-6 k}{12}\bigg)~,
\ee
which is precisely the Virasoro Poisson vertex algebra at central charge $c = -6k$.

\subsection{Classical bulk equivalence}\label{sec:UniversalBulks}
The two boundary presentations above give rise to chiral gravity and Higgsed $\fsl_2$ Chern--Simons theory through the universal bulk construction \cite[Sec.~3.5]{Butson:2016jys}.
This suggests a corresponding equivalence of the bulk theories, which we now establish by an explicit BV quasi-isomorphism.
Explicitly, we consider the map defined by
\be\label{eq:ClassicalIdentification}
	k = -\frac{c}{6} \qquad \begin{aligned}
		\bA^f & = \frac{6}{c}\bmu & \qquad \bA^h & = \frac{1}{2} \partial_z \bmu & \qquad \bA^e & = -\frac{c}{12}\partial_z^2 \bmu + 2\pi \ii \bG \bmu\\
		\bB_f & = \frac{c}{6}\bG & \qquad \bB_h & = 0 & \qquad \bB_e & =\frac{1}{2\pi\ii}\\
	\end{aligned}
\ee
Denote this map of fields by $\Phi$.
It intertwines the BV differentials and pulls back the Chern--Simons BV symplectic form to the gravity BV symplectic form.

We note that the identification $c = -6k$ is totally expected: in the formulation of three-dimensional gravity in terms of an $\SL(2,\R)_k \times \SL(2,\R)_{\bar{k}}$ Chern--Simons theory of \cite{Witten:1988hc}, the levels are identified (in the absence of gravitational Chern--Simons terms) as
\be
	k = \bar{k} = - \frac{\ell}{4G}~,
\ee
where $\Lambda = - \ell^{-2}$ is the cosmological constant and $G$ is Newton's constant, cf. \cite{Coussaert:1995zp}.
Comparing this to the symmetries of asymptotically AdS${}_3$, which is two copies of the Virasoro algebra at central charges $c = \bar{c} = \frac{3\ell}{2G}$ \cite{Brown:1986nw}, we find precisely the above relation.

To formulate the equivalence for these classical field theories, we consider their $P_0$ algebras of classical observables: commutative dg algebras with the degree-one BV Poisson bracket.
We use observables whose Taylor coefficients have smooth, compactly supported first variational derivatives, so that the BV bracket is defined, and complete in total Taylor degree.

For the proof and subsequent Chern--Simons calculations in this section, it is convenient to use
the constant Cartan rescaling
\be
 \widehat{\bA}^{f}=-k\bA^f,\qquad
 \widehat{\bA}^{e}=-k^{-1}\bA^e,\qquad
 \widehat{\bB}_{f}=-k^{-1}\bB_f,\qquad
 \widehat{\bB}_{e}=-k\bB_e,
\ee
leaving the Cartan components unchanged. This preserves the action and BV pairing.
We suppress hats in these coordinates. The map \eqref{eq:ClassicalIdentification} then reads
$\bA^f=\bmu$, $\bB_f=\bG$,
$\bA^e=-\tfrac12\partial^2\bmu+b^{-1}\bG\bmu$, and $\bB_e=b$,
where $b=\ii k/(2\pi)$.
In these coordinates the boundary generators of Section~\ref{sec:ClassicalIdentification} are
$h=(2\pi\ii)B_h$, $f=(-2\pi\ii k)B_f$, and $\psi=kA^e$, where unbold fields denote boundary zero-form components.
On the image of $\Phi$ with $\bmu|=0$, this gives $T=-2\pi\ii G$.

\begin{thm}\label{thm:ClassicalEquivalence}
	Suppose $k\neq 0$ and $c=-6k$, and set $b=\frac{\mathrm{i} k}{2\pi}=-\frac{\mathrm{i} c}{12\pi}$.
	On a bulk coordinate open set $U$, the identification \eqref{eq:ClassicalIdentification} gives a perturbative
	equivalence between chiral gravity and perturbative $\fsl_2$ Chern--Simons theory expanded around the
	vacuum
	\be
	\bA=0,\qquad \bB_h=\bB_f=0,\qquad \bB_e=b.
	\ee
	More precisely, their completed algebras of classical observables are connected by a zigzag of quasi-isomorphisms of $P_0$ algebras
	\be\label{eq:ClassicalPzeroZigzag}
	\mathrm{Obs}^{\mathrm{cl}}_{\mathrm{CS},b}(U)
	\xleftarrow{\ \simeq\ } \mathcal N(U)
	\xrightarrow{\ \simeq\ } \mathrm{Obs}^{\mathrm{cl}}_{\mathrm{grav}}(U).
	\ee
	These maps are compatible with the factorization products, and the underlying equivalence of commutative dg algebras
	is represented by pulling back along the map $\Phi$ as defined in equation \eqref{eq:ClassicalIdentification}.
\end{thm}

\begin{proof}
	Write $B=\mathrm{Obs}^{\mathrm{cl}}_{\mathrm{CS},b}(U)$ and $A=\mathrm{Obs}^{\mathrm{cl}}_{\mathrm{grav}}(U)$.
	Restriction to the image of $\Phi$ defines a surjective morphism of commutative dg algebras $\Phi^*:B\to A$.
	Although $\Phi^*\Omega_{\mathrm{BV},\mathrm{CS}}=\Omega_{\mathrm{BV},\mathrm{grav}}$, this restriction is not itself a Poisson morphism: the image is a symplectic submanifold with additional conjugate directions in the ambient theory.
	We will replace $B$ by a quasi-isomorphic Poisson subalgebra on which restriction does preserve the bracket.
	
	First introduce the following formal coordinates on the Chern--Simons fields:
	\be\label{eq:ClassicalDarbouxCoordinates}
	\begin{aligned}
		x&=\bB_e-b,&\qquad y&=\bB_h,\\
		\widetilde{\bmu}&=\frac{b}{\bB_e}\bA^f,
		&\widetilde{\bG}&=\frac{\bB_e}{b}\bB_f
		-\frac12\partial\bB_h-\frac12\partial^2\bB_e,\\
		\eta&=\bA^h-\frac12\partial\widetilde{\bmu},
		&\xi&=\bA^e-\frac{\bB_f}{b}\widetilde{\bmu}
		+\frac12\partial^2\widetilde{\bmu}.
	\end{aligned}
	\ee
	Here $x,y,\eta,\xi$ are superfields, and $\bB_e^{-1}$ is expanded as a formal power series about $b\neq 0$.
	This change of coordinates is invertible. 
	For compactly supported variations, integration by parts gives
	\be
	\int\big(\bA^f\delta\bB_f+\bA^h\delta\bB_h+\bA^e\delta\bB_e\big)
	=\int\big(\widetilde{\bmu}\delta\widetilde{\bG}+\eta\delta y+\xi\delta x\big).
	\ee
	and so the BV form in these coordinates reads
	\be\label{eq:ClassicalSplitSymplecticForm}
	\Omega_{\mathrm{BV},\mathrm{CS}}
	=\int\big(\delta\widetilde{\bG} \delta\widetilde{\bmu}
	+\delta y \delta\eta+\delta x \delta\xi\big).
	\ee
	The image of $\Phi$ is exactly $x=y=\eta=\xi=0$, where
	$(\widetilde{\bmu},\widetilde{\bG})=(\bmu,\bG)$.
	For $p\geq 1$, let $I^{(p)}$ consist of observables whose Taylor expansion contains at least $p$ factors from the
	normal fields $(x,y,
	\eta, \xi)$.
	In particular, $I=\ker \Phi^* =I^{(1)}$.
	We use this normal Taylor filtration, rather than algebraic powers of an ideal of smeared observables.
	Since $\Phi$ intertwines the differentials, $Q$ is tangent to the slice $\mathrm{im} \Phi$ and preserves every $I^{(p)}$.
	
	Define the Poisson normalizer
	\be
	\mathcal N(U)=\{F\in B:\{F,I\}_{\mathrm{BV}}\subset I\}.
	\ee
	The Leibniz and Jacobi identities make $\mathcal N(U)$ a unital Poisson subalgebra of $B$.
	It is preserved by $Q$, since $Q$ is a derivation of the bracket and preserves $I$.
	By \eqref{eq:ClassicalSplitSymplecticForm}, the normalizer condition says precisely that the Taylor expansion of $F$
	has no term of degree one in the normal fields $(x,y,\eta,\xi)$.
	It follows that, as completed graded vector spaces,
	\be\label{eq:ClassicalNormalizerQuotients}
	\mathcal N(U)=A\oplus I^{(2)},\qquad
	\ker(\Phi^*|_{\mathcal N(U)})=I^{(2)},\qquad
	B/\mathcal N(U)\cong I/I^{(2)}.
	\ee
	In the first equality, $A$ denotes the graded algebra of functions of
	$(\widetilde{\bmu},\widetilde{\bG})$ alone; its inclusion in $B$ need not commute with $Q$.
	The last isomorphism is an isomorphism of complexes, induced by $I\hookrightarrow B$.
	Restriction $\Phi^*|_{\mathcal N(U)}$ is surjective and preserves the Poisson bracket: the normal derivatives of both arguments vanish on the slice, and the remaining bracket is exactly the gravity bracket.
	This constructs the two strict $P_0$ morphisms in \eqref{eq:ClassicalPzeroZigzag}.
	
	It remains to prove that $I^{(2)}$ and $I/I^{(2)}$ are acyclic for the full nonlinear differential.
	Filter by total Taylor degree in all six coordinates of \eqref{eq:ClassicalDarbouxCoordinates}, and let $Q_1$ be the linear part of $Q$ at the vacuum.
	The associated graded differential is determined by
	\be\label{eq:ClassicalNormalDifferential}
	\begin{aligned}
		Q_1\widetilde{\bmu}&=\d'\widetilde{\bmu},
		&Q_1\widetilde{\bG}&=\d'\widetilde{\bG}
		-\frac b2\partial^3\widetilde{\bmu},\\
		Q_1x&=\d'x-2b\eta,
		&Q_1y&=\d'y+2b(\xi+\partial\eta),\\
		Q_1\eta&=\d'\eta,
		&Q_1\xi&=\d'\xi.
	\end{aligned}
	\ee
	The first line is the linear gravity differential, since $-b/2=\ii c/(24\pi)$.
	In the normal sector, replacing $y$ by $y+\partial x$ exhibits two contractible pairs:
	\be
	(Q_1-\d')x=-2b\eta,\qquad
	(Q_1-\d')(y+\partial x)=2b\xi.
	\ee
	More explicitly, define a degree-minus-one homotopy on the internal generators by
	\be
	h(\eta)=-\frac{x}{2b},\qquad
	h(\xi)=\frac{y+\partial x}{2b},\qquad
	h(x)=h(y)=h(\widetilde{\bmu})=h(\widetilde{\bG})=0.
	\ee
	Extend it to differential-form components with the tensor-product Koszul signs and to observables as a graded derivation.
	Then
	\be
	Q_1h+hQ_1=E_{xy \eta \xi},
	\ee
	where $E_{xy\eta \xi}$ counts the number of normal factors.
	Hence $h/n$ contracts the summand of normal degree $n>0$, with arbitrary dependence on the gravity fields.
	This contracts both associated graded complexes in question: $I/I^{(2)}$ has normal degree one and $I^{(2)}$ has normal degree at least two.
	
	Finally, for either $C=I^{(2)}$ or $C=I/I^{(2)}$, let $F^mC$ denote the subcomplex of total Taylor degree at least $m$.
	Since $Q-Q_1$ raises total Taylor degree, the spectral sequence of each finite Taylor quotient $C/F^mC$ has
	\be
	E_1=H^\bullet\big(\operatorname{gr}_F(C/F^mC),Q_1\big)=0.
	\ee
	The induced filtration is finite, so every $C/F^mC$ is acyclic.
	Completeness and the degreewise surjectivity of the truncation maps then imply that $C=\varprojlim_m C/F^mC$ is acyclic as well.
	The exact sequences supplied by \eqref{eq:ClassicalNormalizerQuotients} now show that both arrows in \eqref{eq:ClassicalPzeroZigzag} are quasi-isomorphisms.
	Since $\Phi^*|_{\mathcal N(U)}=\Phi^*\circ(\mathcal N(U)\hookrightarrow B)$, their underlying equivalence agrees with $\Phi^*$.
\end{proof}

All coordinate changes, brackets, and homotopies used here are local differential operations and preserve supports.
The construction is therefore compatible with inclusions of coordinate open sets and with products of observables with disjoint supports.
Thus, we have the following consequence.
\begin{cor}[]
	\label{cor:cs=g_classical}
	The $P_0$-factorization algebras of classical observables of three-dimensional chiral gravity and
	$\mathfrak{sl}_2$
	Chern--Simons theory expanded around the vacuum $\mathbf B_e = b$ are equivalent.
\end{cor}

\subsection{Quantum identification}\label{sec:QuantumBCs}
We now quantize the two bulk-boundary presentations in the holomorphic gauge of Section~\ref{app:PerturbativeQuantization}.
For chiral gravity we compute the Virasoro OPE directly; for Chern--Simons theory we specialize the dg vertex algebra of Section~\ref{sec:QuantumWalgebra} to $\fsl_2$ and pass to cohomology.
Comparing the resulting central charges will determine the parameter matching.
The required Feynman integrals are computed in Appendix~\ref{app:integrals}.

\subsubsection{Three-dimensional chiral gravity and the Virasoro algebra}\label{sec:Virasoro}

We start with the quantization of the gravitational theory described in Section \ref{sec:ClassicalGravity} with the \enquote{Dirichlet} boundary condition that imposes $\bmu| = 0$.
Quantizing the boundary observables of this system will give rise to an explicit deformation quantization of the Virasoro Poisson vertex algebra.
Of course, this deformation quantization will yield the Virasoro VOA; the key computation will be an identification of the quantum correction to the central charge.

The quantum boundary observables are given by normally-ordered products of the zero-form part of $\bG|$, which we denote by $G$, and its holomorphic derivatives; the other derivatives of $G$ are trivialized in cohomology by the 1-form component of $\bG|$.
We compute the OPE of $G$ with itself from two tree-level diagrams and one 1-loop diagram.

The explicit computation of these Feynman integrals is relegated to Section \ref{app:VirasoroOPEs}.
The result is as follows:
\be
\begin{aligned}
	G(z) G(w) & \sim \hbar \left(\frac{1}{(z-w)^4}\frac{-c}{8\pi^2}+\frac{1}{(z-w)^2} \frac{\ii G(w)}{\pi} + \frac{1}{(z-w)} \frac{\ii\partial G(w)}{2 \pi}\right)\\
	& \qquad + \hbar^2 \left(\frac{1}{(z-w)^4}\frac{-13}{8\pi^2}\right)
\end{aligned}
\ee
The terms in the first line arise from tree-level diagrams, whereas the term in the second line is the 1-loop correction.
If we define $T = -2\pi \ii G$, then we see that $T$ is a properly-normalized Virasoro field
\be
	T(z) T(w) \sim \hbar \left(\frac{c_{\hbar}/2}{(z-w)^4}+\frac{2 T(w)}{(z-w)^2} + \frac{\partial T(w)}{z-w}\right)
\ee
at central charge $c_{\hbar} = c + 13 \hbar$.

We can phrase this result as follows.

\begin{thm}\label{thm:1loopVirasoro}
	The holomorphic gauge-fixing of three-dimensional chiral gravity on $\C\times\R_{\geq0}$ with Dirichlet boundary conditions determines a vertex algebra of quantum boundary observables isomorphic to the $\hbar$-adic Virasoro vertex algebra at central charge $c_\hbar=c+13\hbar$.
\end{thm}

The coefficient of the fourth-order pole in the displayed $T(z)T(w)$ OPE gives this shift directly.
In the action normalization above, the same shift is represented by the local chiral quadratic correction
\be
	\frac{\mathrm{i} (13 \hbar)}{48 \pi}\int \d z \bmu \partial_z^3 \bmu~.
\ee

\subsubsection{Higgsed Chern--Simons theory and the $W$-algebra for \texorpdfstring{$\fsl(2)$}{sl2}}\label{sec:WAlgebra}

The chain-level quantum boundary observables are given by normally-ordered products of the zero-form components of the restrictions $\bB_h|$, $\bB_f|$, and $\bA^e|$ together with their holomorphic derivatives; we denote these fields $B_h$, $B_f$, and $A^e$.
The OPEs and quantum-corrected differential can be directly read off from the analysis we performed in Section \ref{sec:Walgebras}.
The OPEs with non-trivial singular terms are given by
\be
\begin{aligned}
	B_h(z) B_h(w) \sim \hbar\left(\frac{1}{(z-w)^2}\frac{-k}{2 \pi^2}\right) &&&& B_h(z) B_f(w) \sim \hbar\left(\frac{1}{z-w}\frac{-2 B_f(w)}{2 \pi \ii}\right)\\
	&&\hspace{-3cm} B_h(z) A^e(w) \sim \hbar \left(\frac{1}{z-w}\frac{-2 A^e(w)}{2 \pi \ii}\right) \hspace{-3cm} &&
\end{aligned}
\ee
with all other OPEs being regular.
As noted before, OPEs of these fields do not receive 1-loop quantum corrections; the quantum-corrections appear in the differential.
The quantum-corrected boundary differential acts on $B_h$ and $A^e$ as
\be
	Q_{\hbar} B_h = 2 \left(\frac{\ii k}{2\pi}\right) A^e \qquad Q_{\hbar} A^e = 0
\ee
and acts on $B_f$ as
\be
\begin{aligned}
	Q_{\hbar} B_f & = -\tfrac{1}{2}(:\!A^e B_h\!: + :\!B_h A^e\!:) + \frac{\ii (k-\hbar)}{2\pi}\partial A^e\\
	& = -:\!A^e B_h\!: + \frac{\ii (k-2\hbar)}{2\pi}\partial A^e
\end{aligned}~.
\ee
If we define
\be
	h = (2\pi \ii) B_h \qquad f = (-2\pi \ii k) B_f \qquad \psi = k A^e
\ee
as in Section \ref{sec:UniversalBulks} then the OPEs take the form
\be
\begin{aligned}
	h(z) h(w) \sim \hbar\left(\frac{2k}{(z-w)^2}\right) &&&& h(z) f(w) \sim \hbar\left(\frac{-2 f(w)}{z-w}\right)\\
	&&\hspace{-2cm} h(z) \psi(w) \sim \hbar \left(\frac{-2 \psi(w)}{z-w}\right) \hspace{-2cm} &&
\end{aligned}
\ee
and the quantum-corrected differential's action is given by
\be
\begin{aligned}
	Q_{\hbar} h = -2 \psi &&&& Q_{\hbar} f = ~ :\!\psi h\!: + (k-2\hbar) \partial\psi\\
	&& Q \psi = 0 &&
\end{aligned}~.
\ee

We note that there is a direct algebraic way to fix this correction.
Keep $Q_{\hbar}h=-2\psi$ and $Q_{\hbar}\psi=0$, which is required by knowing the 1-loop correction to the action is of the form $\int \d z~\bA \partial_z \bA$, and write
$Q_{\hbar}f=:\!\psi h\!:+a\partial\psi$ for $a \in \C[\![\hbar]\!]$, which is the most general expression that respects quantum numbers.
If we require that this is a vertex-algebraic derivation then we are forced to take $a=k-2\hbar$.
For this value, the remaining defining brackets are also preserved and $Q_{\hbar}^2=0$.
This verifies the coefficient directly in the boundary vertex algebra.
We note that this complex appears as the second page of a spectral sequence computing the cohomology of the affine $W$-algebra of $\fsl_2$ at level $k - 2 \hbar$, see e.g. Section 15.2 of \cite{FBZ} for a textbook treatment.

Its cohomology is generated by the field
\be
	T = k^{-1}\left(f + \tfrac{1}{4}:\! h^2 \!: + \tfrac{1}{2}(k-\hbar) \partial h \right)
\ee
which is a Virasoro field at central charge
\be
	c_{W, \hbar} = -6k + 13 \hbar - \frac{6}{k} \hbar^2~.
\ee
The $\hbar^2$ term here arises from the composite, quantum-corrected generator $T$ and its normal ordering; it does not require a two-loop bulk interaction.
The parameters $c$ and $k$ used so far are classical couplings, related by $c=-6k$.
To match the two holomorphic-gauge quantizations, allow the gravity coupling to vary formally as $c(\hbar)=-6(k+\hbar^2/k)$; then $c(\hbar)+13\hbar=c_{W,\hbar}$.
We now explain how the classical equivalence identifies the corresponding spaces of quantizations.

\subsection{Correspondence of quantizations}\label{sec:QuantumEquivalence}

The classical equivalence of BV theories from Corollary \ref{cor:cs=g_classical} extends to a quantum equivalence for
both the bulk three-dimensional theories as well as in the presence of boundary conditions.

\subsubsection{Bulk theories}

The quantum comparison is naturally a correspondence between spaces of quantizations.
Fix the classical parameters $k\neq0$ and $c=-6k$, and disregard field-independent vacuum terms.
At each order in $\hbar$, the space of quantizations
is a torsor for the degree-zero cohomology of its reduced local deformation complex
\cite[Ch.~5, Sec.~11]{Costello:2011renormalization}.
For the cotangent deformations considered here, the space of quantizations of perturbative
Chern--Simons theory is, after choosing a reference quantization,
\be\label{eq:QuantumLevelModuli}
 \hbar H^3(\fsl_2)\llbracket\hbar\rrbracket
 \cong\hbar\C\llbracket\hbar\rrbracket,
\ee
see \cite[Ch.~5, Sec.~14]{Costello:2011renormalization}.

For chiral gravity we find an isomorphic space of cotangent deformations:
\be
 \hbar H^3(\mathfrak w_1)\llbracket\hbar\rrbracket
 \cong\hbar\C\llbracket\hbar\rrbracket.
\ee
Here $\mathfrak w_1$ is the Lie algebra of vector fields on the formal disk, and its cohomology
is continuous Lie algebra, or Gelfand--Fuks, cohomology.
Descent of its degree-three class gives the bulk local functional
$\int\d z\,\bmu\partial_z^3\bmu$
\cite[Thm.~1.7 and Sec.~5.2.1]{Williams:2024local}.
A reference choice identifies the quantization classes of either bulk theory with
$\hbar\C\llbracket\hbar\rrbracket$.
The local construction in the proof of Theorem~\ref{thm:ClassicalEquivalence} identifies the
complexes of local BV deformations of the two theories. Costello's obstruction theory therefore
identifies their spaces of quantizations, including homotopies
\cite[Ch.~5, Sec.~11]{Costello:2011renormalization}.
This gives the following quantum version of Corollary~\ref{cor:cs=g_classical}.

\begin{thm}\label{thm:cs=g_quantum}
 For $k\neq0$ and $c=-6k$, the classical comparison $\Phi$ identifies the spaces of perturbative
 BV quantizations of three-dimensional chiral gravity and $\fsl_2$ Chern--Simons theory expanded
 around the vacuum $\bB_e=b=\mathrm{i} k/(2\pi)$.
 Corresponding quantizations have equivalent $\hbar$-adically complete factorization algebras
 of bulk observables on $\C\times\R$:
 \be\label{eq:QuantumBulkEquivalence}
 \mathrm{Obs}_{\mathrm{CS},b}
 \simeq\mathrm{Obs}_{\mathrm{grav}}.
 \ee
 Modulo $\hbar$, this is the classical equivalence induced by $\Phi$.
\end{thm}

\subsubsection{Quantum equivalence with boundary}

The classical comparison also respects the oper and Dirichlet boundary conditions and identifies
the corresponding local BV deformation complexes.
Rabinovich's boundary obstruction theory then extends the bulk equivalence
\cite[Thm.~4.5.8 and Prop.~4.6.3]{EugeneThesis}.

Horizontal descent for the boundary Dolbeault field $\bG|$ shows that the boundary-supported
local deformation complex, with smooth coefficients, has vanishing $H^0$ and $H^1$.
Thus boundary counterterms introduce no additional quantization parameter.
In the holomorphic-gauge coordinates
\be\label{eq:FormalQuantumLevels}
 k(\hbar)\in k+\hbar\C\llbracket\hbar\rrbracket,\qquad
 c(\hbar)\in c+\hbar\C\llbracket\hbar\rrbracket,
\ee
the boundary calculations in Sections~\ref{sec:Virasoro} and~\ref{sec:WAlgebra} identify the
corresponding classes by equating their Virasoro central charges.
We obtain the following extension of the bulk equivalence.

\begin{thm}\label{thm:cs=g_boundary}
 For $k\neq0$ and $c=-6k$, the bulk comparison extends to an equivalence between the spaces
 of perturbative BV quantizations of Higgsed $\fsl_2$ Chern--Simons theory with oper boundary
 condition and three-dimensional chiral gravity with Dirichlet boundary condition on
 $\C\times\R_{\geq0}$.
 In the holomorphic-gauge coordinates, corresponding quantization classes satisfy
 \be\label{eq:coords}
 c(\hbar)=-6\left(k(\hbar)+\frac{\hbar^2}{k(\hbar)}\right).
 \ee
 Their $\hbar$-adically complete stratified factorization algebras of quantum observables
 are equivalent:
 \be
 \mathrm{Obs}_{\mathrm{CS},\mathrm{oper}}
 \simeq\mathrm{Obs}_{\mathrm{grav},D}.
 \ee
 This equivalence reduces modulo $\hbar$ to the comparison induced by $\Phi$.
\end{thm}

\appendix

\section{Feynman Integrals}
\label{app:integrals}

In this appendix, we compute the relevant Feynman integrals underlying the results in the main text. Specifically, \cref{app:AffineOPEs} deals with the OPEs of boundary observables on the Dirichlet boundary condition in Chern--Simons theory (\cref{sec:AffineVOA}), \cref{app:VirasoroOPEs} with the analogous observables in three-dimensional chiral gravity (\cref{sec:Virasoro}), and \cref{app:EffectiveAction} concerns the 1-loop correction to $\fg$ Chern--Simons theory in the presence of the oper boundary condition described in   \cref{sec:QuantumWalgebra}.

All these computations use Schwinger parameters to express the relevant Feynman integrals. This method is based upon the integral representation of the Euler gamma function \cite{euler_progressionibus_1738}, which can be written as the identity\footnote{The identity itself is of course much older than its application to quantum field theory, which   was perhaps first realized in  the appendix of \cite{schwinger_quantum_1949a}. A related method, often called  Feynman parametrization, is based upon identities such as $\frac 1 {X\cdot Y} = \int_0^1 \frac{\d a}{( X a + Y(1-a))^2}$,  and has been introduced in the appendix of  \cite{feynman_spacetime_1949}. }
\begin{align}\label{Schwinger_trick}
	\frac{1}{X^{\nu}} &= \frac 1 {\Gamma(\nu)} \int_0^\infty \d a \; a^{\nu -1} e^{-a X}.
\end{align}
The key is that, regardless of the parameter $\nu$, the expression $X$ appears linearly in the exponent. 
If $X$ is a quadratic expression in momenta or positions, as is usually the case  for denominators of propagators in Feynman integrals, applying \cref{Schwinger_trick} to each of the propagators yields a combined exponential function whose argument is still quadratic. The spacetime-integration is then Gaussian and can be carried out analytically, and the resulting integral over the parameters $a_e$, one for each edge $e$, is called the Schwinger-parametric form of the integral. 

Schwinger-parametric integrals are by now a standard tool in perturbative QFT. Their structural properties have been studied systematically since the 1950s, e.g.  \cite{nakanishi_general_1957,nakanishi_graph_1971},  covering also questions of convergence, e.g. \cite{brown_angles_2013} or number-theoretic content, e.g. \cite{brown_periods_2010}. Both numerical \cite{bogner_resolution_2008,borowka_numerical_2013,borinsky_tropical_2023a}, as well as symbolic \cite{panzer_algorithms_2015},  integration algorithms are available. As the spacetime integrals after the transformation \cref{Schwinger_trick} are Gaussian, they can in principle always be solved. For the case of scalar integrals, this computation can be done in an abstract general way  for arbitrary Feynman diagrams, and the resulting parametric integrand has as its denominator the Kirchhoff polynomial   
\begin{align}\label{def:Kirchhoff_polynomial}
	\mathcal U(a) = \hspace{-.2cm}\sum_{T \textnormal{ spanning tree }}~ \prod_{e\in T} a_e
\end{align}
or its dual, the first Symanzik polynomial, see e.g.  \cite{bogner_feynman_2010}. 
For non-scalar integrals, no general explicit formulas are available for the parametric representation of arbitrary integrals, although partial progress has been made e.g. in \cite{kreimer_quantization_2013,golz_new_2017}. One notable exception, where indeed an explicit parametric formula is known for arbitrary graphs, are the  Feynman integrals describing anomalies in topological QFT \cite{balduf_combinatorial_2025,balduf_topological_2025}. The present Schwinger-parametric computations arose from a generalization of that previous work to holomorphic theories, although in the present appendix we do not make an effort to derive general formulas for such integrals.

\subsection{Affine VOA OPEs}\label{app:AffineOPEs}
We use the BV pairing and orientation specified in Section~\ref{sec:ClassicalCS}, with the signs fixed by Eq.~\eqref{eq:BVPairingConvention}.
In particular, the holomorphic $K$-frames at different vertices are even.
We normalize the invariant form so that
\begin{align}\label{dual_Coxeter_number}
    f^d_{ac}f^c_{bd}=\Tr_{\fg}(\mathrm{ad}_{T_a}\mathrm{ad}_{T_b})=2h^\vee K_{ab}~.
\end{align}
We compute the three integrals shown in \cref{fig:affineFeynman}, starting with computing the tree-level diagram contributing to the identity component of the OPE.
This term comes from $\hbar$ times the integral
\begin{align} \label{A1_integral1}
	& 2 \times \frac{\ii k K_{ab}}{4 \pi}\int \d w \; \Pi_D(z,0;w,s) \partial_w\Pi_D(0,0;w,s) \nonumber \\
	& \qquad = -\frac{3\ii k K_{ab}}{32 \pi^3} \int \d w \d \ow \d s \; \frac{s \oz \ow}{(s^2 + |z-w|^2)^{\scriptstyle{\frac{3}{2}}}(s^2+|w|^2)^{\scriptstyle{\frac{5}{2}}}}~.
\end{align}
The multiplication of differential forms in the propagators encodes Fermionic properties of the fields in question, and uniquely determines the minus sign in \cref{A1_integral1} upon fixing the convention that the integration measure should be $\d w \d \ow \d s$ (in that order).
In all subsequent integral transformations, we view this measure as positive by definition, and therefore no extra signs will be introduced in any subsequent real transformations or reorderings.%
\footnote{An alternative valid convention would be to treat the integration measure as honest anticommuting differential forms, and keep track of the relative sign in each transformation, but in that case one would have to work with signed domains of integration, which would be equally annoying in terms of notation. To clarify this point, consider the integral $\iint \d x \d y$ over the 2-dimensional plane, which is meant to be positive for positive integrands, and its value does not depend on the order of integration. But if we naively use    $\d x \d y = - \d y \d x$, there would be a spurious sign if we do the $y$-integration first, which needs to be compensated by a sign in the domain of integration if we really want to work with signed surface elements. E.g. in polar coordinates, $r \d r \d \phi = - r \d \phi \d r$ requires the angle to run in the opposite direction in the second case. Instead of dealing with such issues, we use the convention that $\d x \d y = |\d x \d y|$ stays positive under all real transformations, even if we do not write the bars.} %
The factor of two in \cref{A1_integral1} comes from the fact that there are two contractions that give rise to the same integral

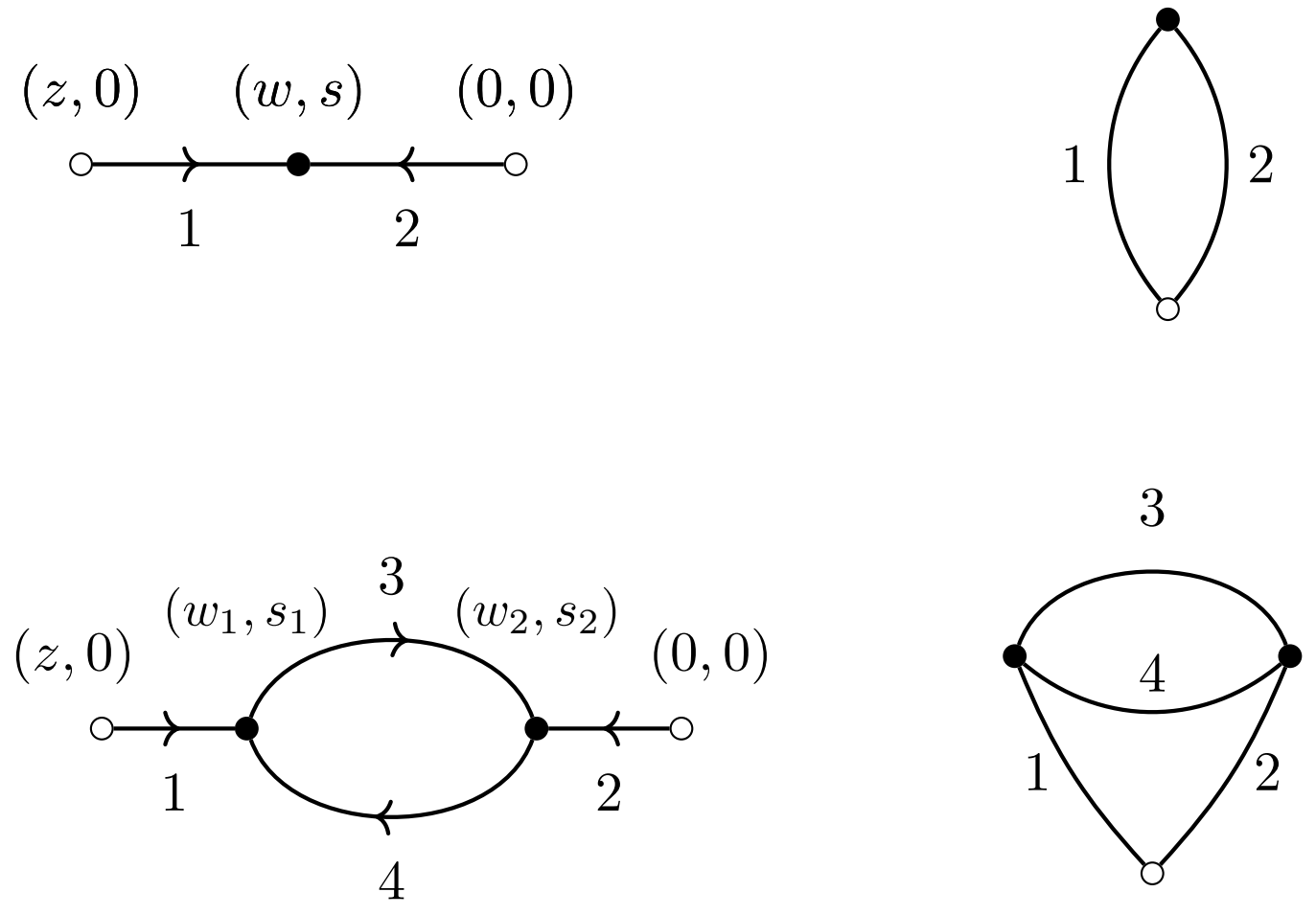
\begin{figure}[ht]
	\centering	\begin{tikzpicture}
		\node[emptyvertex](v1) at (-1.5,0) {};
		\node[emptyvertex] (v2) at (1.5,0) {};
		\node[vertex] (v3) at (0,0) {}; 
		\node (a1) at (-1.5,.5) {$(z,0)$};
		\node (a2) at (0,.5) {$(w,s)$};
		\node (a3) at (1.5,.5) {$(0,0)$};

		\draw[fermion, thick ] (v1) -- (v3) node[midway,below=5pt]{1};
		\draw[fermion, thick] (v2) -- (v3)node[midway,below=5pt]{2}; 
 
		\node[vertex] (v1) at (6,1) {};
		\node[emptyvertex] (v2) at (6,-1) {};
	
	\node (a1) at (-1.5,.5) {$(z,0)$};
	\node (a2) at (0,.5) {$(w,s)$};
	\node (a3) at (1.5,.5) {$(0,0)$}; 
	
	\draw[  thick, bend angle=40,bend left ] (v2) to node[midway, left]{1} (v1) ;
	\draw[ thick,bend angle=40,bend right] (v2) to node[midway, right]{2} (v1); 
	\end{tikzpicture} 
	
	\vspace{1cm}
	
	\begin{tikzpicture}
		\node[vertex] (int1) at (-1,0) {};
		\node[vertex] (int2) at (1,0) {};
		\node[emptyvertex] (b1) at (-2,0) {};
		\node[emptyvertex] (b2) at (2,0) {};
		\node   at (-2.2,.5) {$(z,0)$};
		\node   at (2.2,.5) {$(0,0)$};
		\node   at (-1,.8) {\small $(w_1,s_1)$};
		\node  at (1,.8) {\small $(w_2,s_2)$};
		
		\draw[fermion, thick] (b1) to node[midway, below=5pt]{1}  (int1);
		\draw[fermion, thick] (b2) to   node[midway, below=5pt]{2} (int2);
		\draw[fermion, thick, bend angle=70, bend left] (int1) to  node[midway, above=5pt]{3} (int2);
		\draw[fermion, thick, bend angle=70, bend left] (int2) to node[midway, below=5pt]{4} (int1);

		\node[vertex] (int1) at (4.3,.5) {};
		\node[vertex] (int2) at (6.2,.5) {}; 
		\node[emptyvertex] (b2) at (5.25,-1) {};

		\draw[  thick, bend angle=10,bend left] (b2) to  node[midway, left]{1} (int1);
		\draw[  thick,bend angle=10,bend right] (b2) to  node[midway, right]{2} (int2);
		\draw[ , thick, bend angle=70, bend left] (int1) to node[midway, above=5pt]{3} (int2);
		\draw[ thick, bend angle=40, bend left] (int2) to node[midway, above=0pt]{4} (int1);
	\end{tikzpicture} 
	\caption{\small The Feynman diagrams of \cref{A1_integral1} and \cref{A1_integral3} on the left, and their corresponding completions on the right, which arise by merging all external points. The sum over the spanning trees of the completions defines the Kirchhoff polynomial \cref{def:Kirchhoff_polynomial} which appears in the denominators of parametric integrals.  }
	\label{fig:completions}
\end{figure}

Firstly, we rescale all dimensionful quantities with respect to $z$ by setting $w = \lambda z$, $\ow = \ol{\lambda}\oz$, and $s = |z| \sigma$, this exposes an overall factor $z^{-2}$ and the remaining integral is dimensionless,
\begin{align}
	= -\frac{3\ii k K_{ab}}{32 \pi^3} \frac{1}{z^2} \int \d \lambda \d \ol{\lambda} \d \sigma\; \frac{\sigma \ol{\lambda}}{(\sigma^2 + |1-\lambda|^2)^{\scriptstyle{\frac{3}{2}}}(\sigma^2+|\lambda|^2)^{\scriptstyle{\frac{5}{2}}}}~.
\end{align} 
The domain of integration is still the whole $(\lambda, \ol{\lambda})$ plane and  $\sigma \geq 0$.
Next, we trade the denominators with integrals over additional parametric variables $a_1$, $a_2$ from $0$ to $\infty$ according to \cref{Schwinger_trick},
\be
	-\frac{\ii k K_{ab}}{4 \pi^4} \frac{1}{z^2} \int \d \lambda \d \ol{\lambda} \d \sigma \d^2 a ~\sigma \ol{\lambda} a_1^{\scriptstyle{\frac{1}{2}}} a_2^{\scriptstyle{\frac{3}{2}}} \exp\bigg(-a_1(\sigma^2 + |1-\lambda|^2)-a_2(\sigma^2 + |\lambda|^2)\bigg)~.
\ee
As anticipated, the \enquote{spatial} integral over the $(\lambda, \ol{\lambda})$ plane is now a Gaussian and can be done analytically,%
\footnote{If we spell out the Gaussian in real coordinates $\lambda=x+iy, \ol{\lambda}=x-iy$, this complex-valued transformation in fact gives rise to an overall sign   $\d \lambda \d \ol{\lambda} = -2 i \;d x \d y$.} %
leading to
\begin{align}\label{A1_schwinger1}
	-\frac{k K_{ab}}{2 \pi^3} \frac{1}{z^2} \int \d \sigma \d^2 a \; \frac{\sigma a_1^{\scriptstyle{\frac{3}{2}}} a_2^{\scriptstyle{\frac{3}{2}}}}{\mathcal U^2} \exp\bigg(\frac{a_1^2}{\mathcal U}-a_1 (1+\sigma^2)-a_2\sigma^2\bigg)~.
\end{align}
In the denominator, we recognize the Kirchhoff polynomial (\cref{def:Kirchhoff_polynomial})
\begin{align}\label{A1_integral1_U}
	\mathcal U = a_1 +a_2,
\end{align} 
but of the \emph{completion} of the original Feynman diagram, shown in \cref{fig:completions}.
Likewise, the parameters $\sigma^2$ and $(1+\sigma^2)$ in the exponent formally correspond to what would be the masses of the two propagators, just that in our case they are not constant, but integrated over. 

At this point, the argument of the exponential function  in \cref{A1_schwinger1} is  homogeneous of degree 1 in the Schwinger parameters.%
\footnote{Intuitively, this happens  because the exponent in \cref{Schwinger_trick} is linear in each $a_j$, and the Gaussian integral produces certain homogeneous ratios of the coefficients. A general derivation of these ratios, expressed in terms of graph polynomials,   can be found e.g. in the theses \cite[Sec.~2.1]{panzer_feynman_2015} and  \cite[Sec.~1.3.7]{balduf_dyson_2024}.} %
We introduce a scale variable $\tau$ in the form $a_e=\tau \alpha_e$, this changes the integration measure according to
\begin{align*}
	\d a_1 \d a_2 &= \d (\tau \alpha_1) \d (\tau \alpha_2) =  \d \tau \alpha_1 \tau \d \alpha_2  + \tau\d  \alpha_1 \d \tau \alpha_2+ \tau \d \alpha_1 \tau \d \alpha_2 \\
	&=\tau \d \tau (\alpha_1 \d \alpha_2 - \alpha_2 \d \alpha_1)  + \tau^2 \d \alpha_1 \d \alpha_2 ~.
\end{align*} 
If we choose to perform the integral over $\d \tau$, the last term does not contribute, and we are left with the integration measure $\tau \d \tau\; \omega_\alpha$, where 
\begin{align}\label{projective_measure_alpha}
	\omega_\alpha=\alpha_1\d\alpha_2 - \alpha_2 \d \alpha_1
\end{align}
is a measure on the  projective plane $\mathbb R \mathbb P^1$ in coordinates $[\alpha_1:\alpha_2]$.%
\footnote{Recall that by $\d^2 a$ we mean $|\d a_1 \d a_2|$, so the result should really read $|\tau \d \tau | \times |\omega_\alpha|$. Choosing e.g. the coordinates $\alpha_1 = \cos \phi$ and $\alpha_2=\sin \phi$, the second factor becomes $\omega_\alpha=\d \phi$, and our transformation reproduces the familiar polar coordinates $\d^2 a =r \d r \d \phi$ in the plane, where a coordinate on the positive orthant of $\mathbb R \mathbb P^1$ simply amounts to $\phi \in [0,\frac \pi 2]$.} %
For the scale integral, we use the identity \cref{Schwinger_trick} in reverse direction, 
\begin{align}\label{scale_integral}
	\int_0^\infty \d \tau \; \tau^n e^{-\tau X} &= \frac{1}{X^{n+1}}\Gamma(n+1)~.
\end{align} 
Choosing $n=2$ and $X$  according to \cref{A1_schwinger1}, the result is 
\begin{align}\label{A1_integral_projective}
	-\frac{k K_{ab}}{\pi^3 z^2}\int \d \sigma \; \omega_\alpha  \; \frac{\sigma \alpha_1^{\scriptstyle{\frac{3}{2}}} \alpha_2^{\scriptstyle{\frac{3}{2}}}(\alpha_1 + \alpha_2)}{(\alpha_1 \alpha_2 + \sigma^2 (\alpha_1 + \alpha_2)^2)^3}~.
\end{align}
Up to this point, all integrals were guaranteed to be doable analytically since they were either Gaussian or versions of \cref{scale_integral}.
The remaining integration steps now have to be chosen depending on the precise form of the resulting projective parametric integrand \cref{A1_integral_projective}.
In the present case, noting that $\sigma \d \sigma = \frac 12 \d (\sigma^2)$, the integral over $\sigma$ from $0$ to $\infty$ is particularly easy, and we arrive at the 1-dimensional integral
\begin{align}\label{A1_integral1_last}
	-\frac{k K_{ab}}{4\pi^3 z^2} \int \omega_\alpha \; \alpha_1^{-\frac 12} \alpha_2^{-\frac 12} (\alpha_1+\alpha_2)^{-1}~.
\end{align} 
This last integral can be solved in any chart in $\R \mathbb{P}^1$, such as $\alpha_1=1$, upon which $\omega_\alpha= \d \alpha_2$, and  
\begin{align}\label{A1_integral1_solution}
 -\frac{k K_{ab}}{4\pi^3 z^2} \int_0^\infty \d \alpha_2 \;  \alpha_2^{-\frac 12} (1+\alpha_2)^{-1} = -\frac{k K_{ab}}{4\pi^2 z^2}.
\end{align}

\medskip 

The remaining tree-level contribution to the OPE can be computed similarly.
The integral of interest takes the form
\be
\begin{aligned}
	& f^c_{ab} \int \d w \; \Pi_D(z,0;w,s) \Pi_D(0,0;w,s) \bB_c(w,s)\\
	& \qquad = \frac{f^c_{ab}}{8\pi^2} \int \d w \d \ow \d s \; \frac{s \oz}{(s^2 + |z-w|^2)^{\scriptstyle{\frac{3}{2}}}(s^2+|w|^2)^{\scriptstyle{\frac{3}{2}}}}\bB_c(w,s)\\
	& \qquad = \frac{f^c_{ab}}{8\pi^2} \frac{1}{z} \int \d \lambda \d \ol{\lambda} \d \sigma \;\frac{\sigma}{(\sigma^2 + |1-\lambda|^2)^{\scriptstyle{\frac{3}{2}}}(\sigma^2+|\lambda|^2)^{\scriptstyle{\frac{3}{2}}}}\bB_c(\lambda z, |z|\sigma)~.
\end{aligned}
\ee
To extract the singularities in the OPE of $B_a$ and $B_b$, we must expand $\bB_c$ around $\lambda = \ol{\lambda} = \sigma = 0$; the terms with $\partial_{\ol{\lambda}}$, $\partial_\sigma$ are $Q$-exact and so we can ignore them leading us to
\be
\begin{aligned}
	& \frac{f^c_{ab}}{8\pi^2} \frac{1}{z} \int \d \lambda \d \ol{\lambda} \d \sigma \frac{\sigma}{(\sigma^2 + |1-\lambda|^2)^{\scriptstyle{\frac{3}{2}}}(\sigma^2+|\lambda|^2)^{\scriptstyle{\frac{3}{2}}}}(B_c(0) + \dots),
\end{aligned}
\ee
where the $\dots$ terms are regular.
We can compute these integrals as before: first introduce parametric variables and perform the Gaussian integrals over the $(\lambda, \ol{\lambda})$ plane, then integrate over the scale of the parametric variables, and then perform the projective integrals.
The result is
\be
	\frac{1}{z}\left(\frac{f^c_{ab}}{2\pi \ii}\right) B_c(0) + \dots~.
\ee
\medskip 

We now turn to the 1-loop correction to the OPE shown in \cref{fig:affineFeynman}.
This integral takes the form of $\hbar^2$ times
\be
\begin{aligned}\label{A1_integral3}
	&  2 h^\vee K_{ab} \int \d w_1 \d w_2~\Pi_D(z,0;w_1,s_1) \Pi_D(w_2,s_2;w_1,s_1) \\
	& \hspace{4cm} \times \Pi_D(0,0;w_2,s_2) \Pi_D(w_1,s_1;w_2,s_2)~,
\end{aligned}
\ee
where $h^\vee$ is the dual Coxeter number of $\fg$ according to \cref{dual_Coxeter_number}.
Multiplying the differential forms in the denominators of the propagators, the integral becomes
\be
	-\frac{h^\vee K_{ab}}{z^2}\frac{1}{32 \pi^4}\int\d \lambda_1 \d \ol{\lambda}_1 \d \sigma_1 \d \lambda_2 \d \ol{\lambda}_2\d \sigma_2 \frac{\sigma_1 \sigma_2 (\ol{\lambda}_1 - \ol{\lambda}_2)}{\CD}
\ee
with the denominator
\be
\begin{aligned}
	\CD &= (\sigma_1^2 + |1-\lambda_1|^2)^{\scriptstyle{\frac{3}{2}}}(\sigma_2^2 + |\lambda_2|^2)^{\scriptstyle{\frac{3}{2}}}\\
    &\quad\times ((\sigma_1-\sigma_2)^2+|\lambda_1-\lambda_2|^2)^{\scriptstyle{\frac{3}{2}}} ((\sigma_1+\sigma_2)^2+|\lambda_1-\lambda_2|^2)^{\scriptstyle{\frac{3}{2}}}~.
\end{aligned}
\ee
As before, we introduce parametric variables $a_1$, $a_2$, $a_3$, and $a_4$ for each factor in the denominator and then integrate the Gaussian integrals over the $(\lambda_1, \ol{\lambda}_1)$ and $(\lambda_2, \ol{\lambda}_2)$ planes, leading to the integral
\be
\begin{aligned}\label{A1_1loop_parametric}
	& \frac{h^\vee K_{ab}}{z^2} \frac{2}{\pi^4}\int \d^2 \sigma \d^4 a \;  \frac{\sigma_1 \sigma_2 a_1^{\scriptstyle{\frac{3}{2}}} a_2^{\scriptstyle{\frac{3}{2}}}a_3^{\scriptstyle{\frac{1}{2}}}a_4^{\scriptstyle{\frac{1}{2}}}}{\mathcal U^2} \times\exp\bigg(\frac{a_1^2(a_2+a_3+a_4)}{\mathcal U}\bigg)\\
	& \quad\times\exp\bigg(- a_1 (1+\sigma_1^2) - a_2 \sigma_2^2 - a_3 (\sigma_1 - \sigma_2)^2 - a_4 (\sigma_1 + \sigma_2)^2\bigg)~.
\end{aligned}
\ee
The denominators are again the first Kirchhoff polynomial (\cref{def:Kirchhoff_polynomial}) of the completed graph drawn in the lower row of \cref{fig:completions}, 
\begin{align}\label{A1_1loop_Kirchhoff}
	\mathcal U &= a_1 a_2 + (a_1+a_2)(a_3+a_4)~,
\end{align}
and the argument of the last exponential in \cref{A1_1loop_parametric} has the form $-\sum_e a_e m_e^2$ with what would be the masses of the  propagators. 
Each of the two complex Gaussian integrations has contributed a factor $(-2\pi i)$ to \cref{A1_1loop_parametric}, and $\d^2\sigma\,\d^4a$ denotes the positive real integration measure.
  
We now use \cref{scale_integral} twice to integrate over the scales for both $\sigma$'s and $a$'s, and obtain an integral over a product of projective spaces:
\be
	\frac{h^\vee K_{ab}}{z^2} \frac{1}{\pi^4}\int\omega_\varsigma~\omega_\alpha \frac{\varsigma_1 \varsigma_2\alpha_1^{-\scriptstyle{\frac{1}{2}}}\alpha_2^{-\scriptstyle{\frac{1}{2}}} \alpha_3^{\scriptstyle{\frac{1}{2}}}\alpha_4^{\scriptstyle{\frac{1}{2}}}}{(\alpha_3+\alpha_4)^2(\alpha_1 \varsigma_1^2 + \alpha_2 \varsigma_2^2 + \alpha_3(\varsigma_1 - \varsigma_2)^2 + \alpha_4(\varsigma_1 + \varsigma_2)^2 )^2}~,
\ee
where
\begin{align*}
		\omega_\varsigma &= \varsigma_1 \d \varsigma_2 - \varsigma_2 \d \varsigma_1~,\\
		\omega_\alpha &= \alpha_1 \d \alpha_2\d \alpha_3\d \alpha_4 - \alpha_2 \d \alpha_1\d \alpha_3\d \alpha_4 + \alpha_3\d \alpha_1\d \alpha_2\d \alpha_4 - \alpha_4 \d \alpha_1\d \alpha_2\d \alpha_3~.
\end{align*}
The integrals over $\alpha_1$, $\alpha_2$ are standard, and lead to
\be \label{A1_1loop_projective_rest}
	\frac{h^\vee K_{ab}}{z^2} \frac{1}{\pi^3}\int\omega_\varsigma~\wt{\omega}_\alpha \frac{ \alpha_3^{\scriptstyle{\frac{1}{2}}}\alpha_4^{\scriptstyle{\frac{1}{2}}}}{(\alpha_3+\alpha_4)^2(\alpha_3(\varsigma_1 - \varsigma_2)^2 + \alpha_4(\varsigma_1 + \varsigma_2)^2)}
\ee
with the remaining projective measure
\begin{align}\label{remaining_projective_measure}
		\wt{\omega}_\alpha = \alpha_3 \d \alpha_4 - \alpha_4 \d \alpha_3~.
\end{align}
The two remaining 1-dimensional projective integrals can be solved in many ways, for example by fixing either one of the coordinate to unity like in \cref{A1_integral1_solution}.
To make this computation less repetitive, this time we choose the chart $\alpha_3+\alpha_4=1$, where once again $|\tilde \omega_\alpha|= \d \alpha_4$, and 
\begin{align}
	\int_0^1 d \alpha_4 \;  \frac{ (1-\alpha_4)^{\scriptstyle{\frac{1}{2}}}\alpha_4^{\scriptstyle{\frac{1}{2}}}}{  (1-\alpha_4)(\varsigma_1 - \varsigma_2)^2 + \alpha_4(\varsigma_1 + \varsigma_2)^2 }&=\frac{\pi}{16}\frac{ \varsigma_1^2 + \varsigma_2^2 - (\varsigma_1+\varsigma_2) |\varsigma_1-\varsigma_2| }{\varsigma_1^2 \varsigma_2^2}~.
\end{align}
We do the last integral  in the chart $\varsigma_2=1$ by splitting the domain of integration according to  $\varsigma_1 \gtrless 1$.
The end result of \cref{A1_integral3} is
\begin{align}
	\frac{h^\vee K_{ab}}{16 \pi^2 z^2}\int_0^\infty \d \varsigma_1 \; \frac{ 1+\varsigma_1^2  - (1+\varsigma_1) |\varsigma_1-1| }{  \varsigma_1^2  } &= \frac{h^\vee K_{ab}}{4 \pi^2 z^2}~.
\end{align}

Together with the tree-level terms and $J_a=2\pi\ii B_a$, this yields the level $k-\hbar h^\vee$.
The effective-action computation in Appendix~\ref{app:EffectiveAction} gives the same sign: with all internal fields Dirichlet, its coefficient is
\be
    \frac{\hbar}{2}(2h^\vee K_{ab})I_{DD}
    =\frac{\hbar}{2}(2h^\vee K_{ab})\frac{1}{4\pi\ii}
    =-\frac{\ii\hbar h^\vee K_{ab}}{4\pi}~.
\ee

\subsection{Virasoro OPEs}\label{app:VirasoroOPEs}
We now illustrate how to compute the OPE of the field $G$ generating the boundary observables with itself.
As mentioned in Section \ref{sec:Virasoro}, this OPE has both tree-level and 1-loop contributions; we start with the tree-level terms to show how it reproduces the Virasoro Poisson vertex algebra.
For simplicity, we place one insertion of $G$ at a point $(z,\oz,0)$ away from the origin and the second at the origin $(0,0,0)$.

We first compute the constant term.
This term comes from $\hbar$ times the integral
\be
\begin{aligned}\label{A2_integral1}
	& 2 \times \frac{\ii c}{48 \pi}\int \d w \; \Pi_D(z,0;w,s) \partial_w^3\Pi_D(0,0;w,s)\\
	& \qquad = -\frac{35\ii c}{512 \pi^3} \int \d w \d \ow \d s\; \frac{s \oz \ow^3}{(s^2 + |z-w|^2)^{\scriptstyle{\frac{3}{2}}}(s^2+|w|^2)^{\scriptstyle{\frac{9}{2}}}}\\
	& \qquad = -\frac{35\ii c}{512 \pi^3} \frac{1}{z^4} \int \d \lambda \d \ol{\lambda} \d \sigma \; \frac{\sigma \ol{\lambda}^3}{(\sigma^2 + |1-\lambda|^2)^{\scriptstyle{\frac{3}{2}}}(\sigma^2+|\lambda|^2)^{\scriptstyle{\frac{9}{2}}}}~,
\end{aligned}
\ee
where we have once more set $w = \lambda z$, $\ow = \ol{\lambda}\oz$, and $s = |z| \sigma$; the factor of two comes from the fact that there are two contractions that give rise to the same integral.
Introducing Schwinger parameters according to \cref{Schwinger_trick}, we get
\be
	-\frac{\ii c}{48 \pi^4} \frac{1}{z^4} \int \d \lambda \d \ol{\lambda} \d \sigma \d^2 a ~\sigma \ol{\lambda}^3 a_1^{\scriptstyle{\frac{7}{2}}} a_2^{\scriptstyle{\frac{1}{2}}} \exp\bigg(-a_1(\sigma^2 + |\lambda|^2)-a_2(\sigma^2 + |1-\lambda|^2)\bigg)~.
\ee
The Gaussian integral over $(\lambda,\ol{\lambda})$ produces an extra factor $(-2i)$ as before, and we find the parametric integral
\be
	-\frac{c}{24 \pi^3} \frac{1}{z^4} \int \d \sigma \d^2 a \;  \frac{\sigma a_1^{\scriptstyle{\frac{7}{2}}} a_2^{\scriptstyle{\frac{7}{2}}}}{\mathcal U^4} \exp\left(\frac{a_2^2}{\mathcal U} - a_1 \sigma^2 - a_2(\sigma^2 +1)\right)~.
\ee
The denominators   are the same $\mathcal U=\alpha_1+\alpha_2$ as in \cref{A1_integral1_U} because the topology of the diagrams is the same, shown in the upper row of \cref{fig:completions}.
The different overall power of $\mathcal U$ reflects the different powers of propagators in \cref{A2_integral1} compared to \cref{A1_integral1}. 

The integral over the joint scale $\tau$ of the parameters $a_j$ is completely analogous to above, using \cref{scale_integral} we find
\be
	-\frac{c}{\pi^3 z^4} \int \d \sigma  \omega_\alpha \;\frac{\sigma \alpha_1^{\scriptstyle{\frac{7}{2}}} \alpha_2^{\scriptstyle{\frac{7}{2}}}(\alpha_1 + \alpha_2)}{(\alpha_1 \alpha_2 + \sigma^2 (\alpha_1 + \alpha_2)^2)^5}
\ee
with the projective measure $\omega_\alpha$ of \cref{projective_measure_alpha}. 
The $\sigma$-integral yields
\begin{align}
	-\frac{c}{8 \pi^3 z^4} \int \omega_\alpha \; \alpha_1^{-\frac 12} \alpha_2^{-\frac 12} (\alpha_1+\alpha_2)^{-1}.
\end{align}
Up to the prefactor, this is identical with \cref{A1_integral1_last}, and the end result is 
\be
	-\frac{c}{8\pi^3 z^4} \int_0^\infty \d \alpha_2 \;  \alpha_2^{-\frac 12} (1+\alpha_2)^{-1} = \frac{1}{z^4}\frac{-c}{8\pi^2}~.
\ee 

In the remaining tree-level contribution to the OPE, the two contractions of the commuting physical field components add, so the integral of interest takes the form
\be
\begin{aligned}
	& \int \d w \; \bigg(\Pi_D(z,0;w,s) \partial_w \Pi_D(0,0;w,s)\\
    &\qquad  \qquad  + \Pi_D(0,0;w,s) \partial_w \Pi_D(z,0;w,s) \bigg) \bG(w,s)\\
	& \qquad = -\frac{3}{16\pi^2} \int \d w \d \ow \d s \; \bigg(\frac{s \oz \ow}{(s^2 + |z-w|^2)^{\scriptstyle{\frac{3}{2}}}(s^2+|w|^2)^{\scriptstyle{\frac{5}{2}}}}\\
	& \hspace{4cm} +\frac{s \oz (\oz-\ow)}{(s^2 + |z-w|^2)^{\scriptstyle{\frac{5}{2}}}(s^2+|w|^2)^{\scriptstyle{\frac{3}{2}}}}\bigg)\bG(w,s)
\end{aligned}
\ee

\begin{align*}
	& =	 -\frac{3}{16\pi^2} \frac{1}{z^2} \int \d \lambda \d \ol{\lambda} \d \sigma \;\bigg(\frac{\sigma \ol{\lambda}}{(\sigma^2 + |1-\lambda|^2)^{\scriptstyle{\frac{3}{2}}}(\sigma^2+|\lambda|^2)^{\scriptstyle{\frac{5}{2}}}}\\
	& \hspace{2.8cm} +\frac{\sigma (1-\ol{\lambda})}{(\sigma^2 + |1-\lambda|^2)^{\scriptstyle{\frac{5}{2}}}(\sigma^2+|\lambda|^2)^{\scriptstyle{\frac{3}{2}}}}\bigg)(G(0) + z \lambda \partial G(0) + \dots)~.
\end{align*}
In the last step, in order to extract the singularities in the OPE of $G$ with itself, we have expanded $\bG$ around $\lambda = \ol{\lambda} = \sigma = 0$, and left out the terms with $\partial_{\ol{\lambda}}$, $\partial_\sigma$, which are $Q$-exact.
The integrals are then computed by the same sequence of steps as before, the end result is
\be
	\frac{\ii}{\pi z^2}G(0) + \frac{\ii}{2\pi z}\partial G(0)~.
\ee
Together with above term, we are led to precisely the $\hbar^1$ term in the OPE presented in Section \ref{sec:Virasoro}.

\medskip 

The 1-loop correction to the OPE is given by a sum of four terms:
\be
\begin{aligned}\label{A2_1loop_integrals}
	&(A) \qquad  \int \d w_1 \d w_2~\Pi_D(z,0;w_1,s_1) \partial_{w_1} \Pi_D(w_2,s_2;w_1,s_1)\\
	& \hspace{3cm} \times \Pi_D(0,0;w_2,s_2) \partial_{w_2} \Pi_D(w_1,s_1;w_2,s_2)\\
	&(B) \qquad  +\int \d w_1 \d w_2~\Pi_D(z,0;w_1,s_1) \partial_{w_1} \Pi_D(w_2,s_2;w_1,s_1)\\
	& \hspace{3cm} \times  \Pi_D(w_1,s_1;w_2,s_2) \partial_{w_2} \Pi_D(0,0;w_2,s_2)\\
	&(C) \qquad  +\int \d w_1 \d w_2~\Pi_D(w_2,s_2;w_1,s_1) \partial_{w_1} \Pi_D(z,0;w_1,s_1)\\
	& \hspace{3cm} \times  \Pi_D(0,0;w_2,s_2) \partial_{w_2} \Pi_D(w_1,s_1;w_2,s_2)\\
	&(D) \qquad  +\int \d w_1 \d w_2~\Pi_D(w_2,s_2;w_1,s_1) \partial_{w_1} \Pi_D(z,0;w_1,s_1)\\
	& \hspace{3cm} \times  \Pi_D(w_1,s_1;w_2,s_2) \partial_{w_2} \Pi_D(0,0;w_2,s_2)~.
\end{aligned}
\ee
The first step, again, is to multiply out the differential forms and rescale with respect to $z$.
All four integrals take the shape
\begin{align}\label{A2_denominator}
		&\frac{9}{256 \pi^4}\frac{1}{z^4}\int\d \lambda_1 \d \ol{\lambda}_1 \d \sigma_1 \d \lambda_2 \d \ol{\lambda}_2\d \sigma_2 \; \frac{\CN_j}{\CD_{\nu_1,\nu_2,\nu_3,\nu_4}}~, \\
	 \text{where}\qquad    &\CD_{\nu_1,\nu_2,\nu_3,\nu_4}  = (\sigma_1^2 + |1-\lambda_1|^2)^{\nu_1}(\sigma_2^2 + |\lambda_2|^2)^{\nu_2} \nonumber \\
	&\qquad\times ((\sigma_1-\sigma_2)^2+|\lambda_1-\lambda_2|^2)^{\nu_3} ((\sigma_1+\sigma_2)^2+|\lambda_1-\lambda_2|^2)^{\nu_4}~. \nonumber
\end{align}
\begin{equation}
\begin{aligned}
	\CN_A &= \sigma_1 \sigma_2 (\ol \lambda_1 - \ol \lambda_2)^3~, \quad &&\CD_{\frac 3 2, \frac 3 2, \frac 5 2 , \frac 5 2}~, \\
	\CN_B &= \sigma_1 \sigma_2(\ol \lambda_1 - \ol \lambda_2)^2 \ol \lambda_2  \big( \sigma_1^2 + \sigma_2^2+|\lambda_1-\lambda_2|^{2}   \big)~, \quad &&\CD_{\frac 3 2, \frac 5 2, \frac 5 2, \frac 5 2}~, \\
	\CN_C &= \sigma_1 \sigma_2(\ol \lambda_1 - \ol \lambda_2)^2   (1-\ol \lambda_1)  \big( \sigma_1^2 + \sigma_2^2+|\lambda_1-\lambda_2|^{2}   \big)~, \quad &&\CD_{\frac 5 2, \frac 3 2, \frac 5 2, \frac 5 2}~, \\
	\CN_D &= \sigma_1 \sigma_2(\ol \lambda_1 - \ol \lambda_2)   (1-\ol \lambda_1) \ol \lambda_2  ~ , \quad &&\CD_{\frac 5 2, \frac 5 2, \frac 3 2, \frac 3 2}~.
\end{aligned}
\end{equation}
We observe that $B$ and $C$ are identical upon exchanging $\bar \lambda_1 \leftrightarrow 1- \bar \lambda_2$, which does not alter the value of the integral over the full $(\lambda,\bar \lambda)$ plane, hence it suffices to compute $A,B$, and $D$.  

As before, we introduce parametric variables $a_1$, $a_2$, $a_3$, and $a_4$ for each factor in the denominator, following the numbering convention introduced in the denominator of \cref{A2_denominator}.
The results involve the  Kirchhoff polynomial $\mathcal U = a_1 a_2  + (a_1+a_2)( a_3+a_4)$ from \cref{A1_1loop_Kirchhoff}, since all integrals arise from the isomorphic 1-loop graph shown in \cref{fig:completions}.
One finds
\begin{align*}
&(A) &	&-\frac{1}{\pi^4 z^4} \int \d^2 \sigma \d^4 a \; \frac{\sigma_1 \sigma_2 a_1^{\frac 7 2} a_2^{\frac 7 2}  a_3^{\frac 3 2} a_4^{\frac 32} }{\mathcal U^4} \exp \left( -\frac{a_1 a_2(a_3+a_4)}{\mathcal U} \right) \\
	&&&\hspace{2cm} \times \exp \left(  -a_1 \sigma_1^2 - a_2 \sigma_2^2 - a_3 (\sigma_1-\sigma_2)^2 -a_4 (\sigma_1+\sigma_2)^2  \right)~, \\
	&(B)&	&-\frac{2}{3\pi^4 z^4} \int \d^2 \sigma \d^4 a \; \frac{\sigma_1 \sigma_2 a_1^{\frac 7 2} a_2^{\frac 7 2}  a_3^{\frac 3 2} a_4^{\frac 32} \cdot N_B}{\mathcal U^6} \exp \left( -\frac{a_1 a_2(a_3+a_4)}{\mathcal U} \right) \\
	&&&\hspace{2cm}  \times \exp \left(  -a_1 \sigma_1^2 - a_2 \sigma_2^2 - a_3 (\sigma_1-\sigma_2)^2 -a_4 (\sigma_1+\sigma_2)^2  \right)~ , \\
	 &(D) &	&-\frac{1}{\pi^4 z^4}\int \d^2 \sigma \d^4 a \; \frac{\sigma_1 \sigma_2 a_1^{\frac 72} a_2^{\frac 72} a_3^{\frac 12} a_4^{\frac 12} (a_3+a_4)^2}{\mathcal U^4} \exp \left( \frac{a_1^2(a_2+a_3+a_4)}{\mathcal U} \right) \\
	&&	&\hspace{2cm}  \times \exp \left( -a_1(1+\sigma_1^2) - a_2 \sigma_2^2 - a_3(\sigma_1-\sigma_2)^2 - a_4(\sigma_1+\sigma_2)^2  \right)~ ,
\end{align*} 
where the numerator of $B$ is
\begin{align*}
	N_B &= (\sigma_1^2+\sigma_2^2)(a_3+a_4) \mathcal U^2  + 2 a_1 a_2 (a_3+a_4)(a_2+3(a_3+a_4))\\
	&\quad + 3 a_2^2(a_3+a_4)^2 +a_1^2(a_2^2(a_3+a_4-1)+2 a_2(a_3+a_4)+3(a_3+a_4)^2)~.
\end{align*} 
Next, we make these integrals projective by transforming $\d^2 \sigma \rightarrow \tau \d \tau \omega_\varsigma$, and $\d^4 a \rightarrow \tau^3 \d \tau \omega_a$, respectively, where now $\omega_a$ is a positive measure on $\mathbb{R}\mathbb{P}^3$ in the projective coordinates $[\alpha_1:\alpha_2:\alpha_3:\alpha_4]$.
This produces 
\be
\begin{aligned}
	(A) && &-\frac{3}{\pi^4 z^4} \int \omega_\varsigma \omega_\alpha \; \frac{\varsigma_1 \varsigma_2 \alpha_1^{-\frac 12} \alpha_2^{-\frac12} \alpha_3^{\frac 32} \alpha_4^{\frac 3 2} }{(\alpha_3+\alpha_4)^4 \times y^2 }~,\\
(B) && &	-\frac{2}{\pi^4 z^4} \int \omega_\varsigma \omega_\alpha  \;\frac{\varsigma_1 \varsigma_2 \alpha_1^{-\frac 1 2} \alpha_2^{-\frac 12}\alpha_3^{\frac 32} \alpha_4^{\frac 32} \times x_B }
	{(\alpha_3+\alpha_4)^4 \times y^3}~,\\
(D) &&&	-\frac{3}{\pi^4 z^4} \int \omega_\varsigma \omega_\alpha \; \frac{\varsigma_1 \varsigma_2 \alpha_1^{-\frac 12} \alpha_2^{-\frac 12} \alpha_3^{\frac 12} \alpha_4^{\frac 12} }{(\alpha_3+\alpha_4)^2 \times y^2} 
\end{aligned} 
\ee
with the shorthands
\begin{align*}
	x_B &= 3 \varsigma_1^2 \alpha_1 + 3 \varsigma_2^2 \alpha_2   + 5 (\varsigma_1^2+\varsigma_2^2)(\alpha_3+\alpha_4) -6\varsigma_1\varsigma_2(\alpha_3-\alpha_4) ~,\\
	y&=  (\alpha_1+\alpha_3+\alpha_4)\varsigma_1^2 - 2 (\alpha_3-\alpha_4)\varsigma_1 \varsigma_2 + (\alpha_2 + \alpha_3 +\alpha_4)\varsigma_2^2~ .
\end{align*}
After integrating both $\alpha_1$ and $\alpha_2$ from $0$ to $\infty$, we obtain  
\begin{align*}
 (A) &&&	- \frac{3}{\pi^3 z^4} \int \omega_\varsigma 	\wt{\omega}_\alpha\; \frac{\alpha_3^{\frac 32} \alpha_4^{\frac 32}}{(\alpha_3+\alpha_4)^4 \big( \alpha_3(\varsigma_1-\varsigma_2)^2 + \alpha_4(\varsigma_1+\varsigma_2)^2\big)}~, \\
 (B) &&&	-\frac{4}{\pi^3 z^4} \int \omega_\varsigma 	\wt{\omega}_\alpha \; \frac{\alpha_3^{\frac 3 2} \alpha_4^{\frac 32} \big(2(\alpha_3+\alpha_4)(\varsigma_1^2 + \varsigma_2^2) -3(\alpha_3-\alpha_4)\varsigma_1 \varsigma_2 \big) }{(\alpha_3+\alpha_4)^4 \big( \alpha_3 (\varsigma_1-\varsigma_2)^2 + \alpha_4(\varsigma_1+\varsigma_2)^2\big)^2}~, \\ 
 (D) &&&	-\frac{3}{\pi^3 z^4} \int \omega_\varsigma 	\wt{\omega}_\alpha \; \frac{\alpha_3^{\frac 12} \alpha_4^{\frac 12}}{(\alpha_3+\alpha_4)^2 \big( \alpha_3(\varsigma_1-\varsigma_2)^2 + \alpha_4 (\varsigma_1+\varsigma_2)^2 \big) }~.
\end{align*} 
Here, $\wt{\omega}_\alpha$ is the 2-dimensional projective measure \cref{remaining_projective_measure}, and we notice that the integral in $(D)$ coincides with \cref{A1_1loop_projective_rest}.
All three integrals simplify in the chart $\alpha_3+\alpha_4=1$, and evaluate to
\begin{align*}
	(A) && \frac{3}{256\pi^2 z^4} \int \omega_\varsigma \frac{(\varsigma_1^2+\varsigma_2^2)(\varsigma_1^4 - 4 \varsigma_1^2 \varsigma_2^2 + \varsigma_2^4) - (\varsigma_1+\varsigma_2)^3 |\varsigma_1-\varsigma_2|^3}{\varsigma_1^4 \varsigma_2^4} &=-\frac{1}{8\pi^2 z^4}~, \\
 (B) & 	&-\frac{3}{32 \pi^2 z^4} \int \omega_\varsigma \frac{\varsigma_1^2 + \varsigma_2^2 - (\varsigma_1+\varsigma_2) |\varsigma_1-\varsigma_2|}{\varsigma_1^2 \varsigma_2^2}  &= -\frac{3}{8\pi^2 z^4}~, \\
 (D) &  & -\frac{3}{16 \pi^2 z^4} \int \omega_\varsigma \; \frac{\varsigma_1^2 + \varsigma_2^2 - (\varsigma_1+\varsigma_2)|\varsigma_1-\varsigma_2|}{\varsigma_1^2 \varsigma_2^2} &= -\frac{3}{4\pi^2 z^4}~.
\end{align*}
The final result of the 1-loop integral \cref{A2_1loop_integrals} therefore is 
\begin{align*}
	(A) + 2\times (B) + (D) &= -\frac{13}{8\pi^2} \frac{1}{z^4}~.
\end{align*}

\subsection{Effective Action}\label{app:EffectiveAction}

\begin{figure}[ht]
	\centering
	\begin{tikzpicture}
		\node (int1) at (-0.75,0) {};
		\node (int2) at (0.75,0) {};
		\node (b1) at (-1.5,0) {};
		\node (b2) at (1.5,0) {};
		\node (a1) at (-2,0) {$\bA^h$};
		\node (a2) at (2,0) {$\bA^h$};
		\node (e1) at (0,1) {$\bA^e/\bB_e$};
		\node (e2) at (0,-1) {$\bA^e/\bB_e$};
		
		\draw[fermion, thick] (b1.center) to [in=180, out=0] (int1.center);
		\draw[fermion, thick] (b2.center) to [in=0, out=180] (int2.center);
		\draw[fermion, thick] (int1.center) to [in=90, out=90] (int2.center);
		\draw[fermion, thick] (int2.center) to [in=270, out=270] (int1.center);
	\end{tikzpicture} \hspace{1cm} \begin{tikzpicture}
		\node (int1) at (-0.75,0) {};
		\node (int2) at (0.75,0) {};
		\node (b1) at (-1.5,0) {};
		\node (b2) at (1.5,0) {};
		\node (a1) at (-2,0) {$\bA^h$};
		\node (a2) at (2,0) {$\bA^h$};
		\node (e1) at (0,1) {$\bA^f/\bB_f$};
		\node (e2) at (0,-1) {$\bA^f/\bB_f$};
		
		\draw[fermion, thick] (b1.center) to [in=180, out=0] (int1.center);
		\draw[fermion, thick] (b2.center) to [in=0, out=180] (int2.center);
		\draw[fermion, thick] (int1.center) to [in=90, out=90] (int2.center);
		\draw[fermion, thick] (int2.center) to [in=270, out=270] (int1.center);
	\end{tikzpicture}
	\caption{Two-vertex wheel diagrams contributing to $k_{hh}$ in the effective action.}
	\label{fig:khh}
\end{figure}

\begin{figure}[ht]
	\centering
	\begin{tikzpicture}
		\node (int1) at (-0.75,0) {};
		\node (int2) at (0.75,0) {};
		\node (b1) at (-1.5,0) {};
		\node (b2) at (1.5,0) {};
		\node (a1) at (-2,0) {$\bA^e$};
		\node (a2) at (2,0) {$\bA^f$};
		\node (e1) at (0,1) {$\bA^e/\bB_e$};
		\node (e2) at (0,-1) {$\bA^h/\bB_h$};
		
		\draw[fermion, thick] (b1.center) to [in=180, out=0] (int1.center);
		\draw[fermion, thick] (b2.center) to [in=0, out=180] (int2.center);
		\draw[fermion, thick] (int1.center) to [in=90, out=90] (int2.center);
		\draw[fermion, thick] (int2.center) to [in=270, out=270] (int1.center);
	\end{tikzpicture} \hspace{1cm} \begin{tikzpicture}
		\node (int1) at (-0.75,0) {};
		\node (int2) at (0.75,0) {};
		\node (b1) at (-1.5,0) {};
		\node (b2) at (1.5,0) {};
		\node (a1) at (-2,0) {$\bA^e$};
		\node (a2) at (2,0) {$\bA^f$};
		\node (e1) at (0,1) {$\bA^h/\bB_h$};
		\node (e2) at (0,-1) {$\bA^f/\bB_f$};
		
		\draw[fermion, thick] (b1.center) to [in=180, out=0] (int1.center);
		\draw[fermion, thick] (b2.center) to [in=0, out=180] (int2.center);
		\draw[fermion, thick] (int1.center) to [in=90, out=90] (int2.center);
		\draw[fermion, thick] (int2.center) to [in=270, out=270] (int1.center);
	\end{tikzpicture}
	\caption{Two-vertex wheel diagrams contributing to $k_{ef}$ in the effective action.}
	\label{fig:kef}
\end{figure}

We evaluate the two-vertex wheel  graphs shown in \cref{fig:khh,fig:kef}, which enter Theorem~\ref{thm:oper1loop} for $\fg$ Chern--Simons theory with the oper boundary condition of Section~\ref{sec:QuantumWalgebra}.
The local chiral quadratic vertex is
\be\label{eq:OperOneLoopLocalVertex}
 I^{(1)}_{\mathrm{loc},2}
 =\frac{\ii\hbar}{4\pi}\int\d z~B(\bA,\partial_z\bA)~.
\ee
We first compute the analytic integral for each pair of boundary conditions and then insert the Lie-algebraic factors.

Each of the two-vertex wheel graphs leads to an integral of the form
\be
\begin{aligned}
	& \hbar \sum_{c,d} \tfrac{1}{2} f^c_{ad} f^d_{bc} \int \d z \d w \; \bA^a (z,\oz,t) \Pi_{\CB_d}(w, s;z, t)  \bA^b(w,\ow,s) \Pi_{\CB_c}(z, t;w, s),
\end{aligned}
\ee
where $\CB_c,\CB_d\in\{D,N\}$ specify the boundary conditions on the corresponding $\bA$ components.
As above, we do not explicitly write the dependence of propagators on antiholomorphic coordinates, $\Pi(w,s;z,t) = \Pi(w,\ow,s; z,\oz,t)$. 
To extract \eqref{eq:OperOneLoopLocalVertex}, take the two-form component of $\bA^a(z,\oz,t)$ and the zero-form component of $\bA^b(w,\ow,s)$, retaining only holomorphic derivatives in the Taylor expansion of the latter.
The relevant kernel is proportional to $\ow-\oz$ times a function of $|w-z|^2,t,s$, so angular integration selects precisely $(w-z)\partial_z\bA^b(z,\oz,t)$.
The resulting expression is
\be
	\hbar \tfrac{1}{2} \sum_{c,d} f^c_{ad} f^d_{bc} \int \d z ~ \bA^a (z,\oz,t) \partial_z \bA^b(z,\ol{z},t) I_{\CB_c \CB_d}(z,\oz,t)
\ee
where
\be
	I_{\CB_c \CB_d}(z,\oz,t) = \int \d w (w-z) \Pi_{\CB_c}(z, t;w, s) \Pi_{\CB_d}(w, s;z,t) ~.
\ee
Of the three cases  $DD$, $NN$, and mixed boundary conditions,
the latter do not contribute since $\Pi_D(z, t;w, s) \Pi_N(w, s;z, t) = 0$.
Furthermore, $\Pi_{D}(z, t;w, s) \Pi_{D}(w, s;z, t)=-\Pi_{N}(z, t;w, s) \Pi_{N}(w, s;z, t)$, it therefore remains to evaluate only one non-trivial integral, 
\be
\begin{aligned}
	I_{DD} &= \int \d w \; (w-z) \Pi_{D}(z, t;w, s) \Pi_{D}(w, s;z, t) = - I_{NN}\\
	& = \frac{t}{8\pi^2} \int \d w \d \ow \d s \; \frac{|w-z|^2}{((t-s)^2 + |z-w|^2)^{\scriptstyle{\frac{3}{2}}} ((t+s)^2 + |z-w|^2)^{\scriptstyle{\frac{3}{2}}}} 
\end{aligned}
\ee
$I_{DD}$ is overall homogeneous of degree zero in dimensionful quantities $(w,\ol{w},s,t)$, hence upon rescaling $w\rightarrow \lambda z, \ow \rightarrow \ol \lambda \ol z, s \rightarrow |z| \sigma, t \rightarrow |z| \tau$, it is  independent of $z, \oz$:
\begin{align}
	-\frac{i \tau }{\pi^2}\int \d^2 a \d \sigma \; \frac{a_1^{\frac 12} a_2^{\frac 12}}{(a_1+ a_2)^2} \exp \Big( -a_1 (\sigma-\tau)^2 - a_2 (\sigma+\tau)^2 \Big). 
\end{align}
This integral can be solved by the same sequence of steps as all the previous ones, the end result is independent of $\tau$, 
\begin{align}
-I_{NN}=	I_{DD} &= -\frac{i}{4\pi}=\frac 1 {4\pi i}, \qquad I_{DN}=I_{ND}=0.
\end{align}

\medskip 
Knowing the value of the integrals, we can reproduce the assertions of \cref{thm:oper1loop}.
For the oper boundary condition, $\bA$ is Neumann in $\fn_+$ and Dirichlet in $\mathfrak b_-$.
Inserting these integrals into the color sum gives
\be
	\frac{\ii \hbar}{4\pi}B(T_a, T_b) = \frac{\hbar}{2}\left(\frac{1}{4\pi \ii} f_{a \ol{b}}^{\ol a} f_{b \ol{a}}^{\ol b} - \frac{1}{4\pi \ii} f_{a \beta}^{\alpha} f_{b \alpha}^{\beta}\right),
\ee 
which reproduces the bilinear form in the proof of  \cref{thm:oper1loop}.

For $\fg=\fsl_2$, put $k_{hh}=B(h,h)$ and $k_{ef}=B(e,f)$.
The two diagrams in Figure~\ref{fig:khh} have equal color factors and opposite analytic integrals, so $k_{hh}=0$.

For $k_{ef}$, the left diagram in Figure~\ref{fig:kef} has mixed boundary conditions and vanishes; the right diagram has two Dirichlet propagators and color factor $2$.
Including the symmetry factor $1/2$ gives
\be
 k_{ef}=-4\pi\ii\,\big(\tfrac12\cdot2\big)I_{DD}=-1~.
\ee
Thus the explicit $\fsl_2$ calculation agrees with  \cref{thm:oper1loop}.

\clearpage
\bibliography{CSquant}
\bibliographystyle{amsalpha}
	
\end{document}